\documentclass[ALICE,manyauthors]{cernphprep}

\newcommand{\Kzs}{\rm K^{ 0}_S}
\newcommand{\Kpm}{\rm K^{\pm}}
\newcommand{\Kz}{\rm K^0}
\newcommand{\Kzb}{\rm \overline{K}\,^0}
\newcommand{\Kplus}{\rm K^+}
\newcommand{\Kmin}{\rm K^-}
\newcommand{\Kaon}{\rm K}
\newcommand{\fz}{\rm f_{\rm 0}}
\newcommand{\az}{\rm a_{\rm 0}}
\newcommand{\pro}{\rm p}
\newcommand{\pbar}{\rm \overline{p}}
\usepackage[comma,square,numbers,sort&compress]{natbib}
\usepackage{hyperref}
\usepackage{lineno}
\usepackage{amsmath}
\usepackage{bm}

\begin{document}%

\begin{titlepage}
\PHyear{2015}
\PHnumber{106}      
\PHdate{23 Apr}  
%

\title{One-dimensional pion, kaon, and proton femtoscopy \\ in Pb--Pb collisions at $\mathbf{\sqrt{\bm{s}_{\mathrm {\bf NN}}}=2.76}$~TeV}
\ShortTitle{One-dimensional femtoscopy in Pb--Pb collisions}   

\Collaboration{ALICE Collaboration\thanks{See Appendix~\ref{app:collab} for the list of collaboration members}}
\ShortAuthor{ALICE Collaboration} 

\begin{abstract}
The size of the particle emission region in high-energy collisions can be deduced using the femtoscopic correlations of particle pairs at low relative momentum. Such correlations arise due to quantum statistics and Coulomb and strong final state interactions. In this paper, results are presented from femtoscopic analyses of $\pi^{\pm}\pi^{\pm}$, $\Kpm\Kpm$, $\Kzs\Kzs$, $\pro\pro$, and $\pbar\pbar$  correlations from Pb-Pb collisions at $\sqrt{s_{\mathrm {NN}}}=2.76$~TeV by the ALICE experiment at the LHC. One-dimensional radii of the system are extracted from  correlation functions in terms of the invariant momentum difference of the pair. The comparison of the measured radii with the predictions from a hydrokinetic model is discussed. The pion and kaon source radii display a monotonic decrease with increasing  average pair transverse mass $m_{\rm T}$  which is consistent with hydrodynamic model predictions for central collisions. The kaon and proton source sizes can be reasonably described by approximate $m_{\rm T}$-scaling.
\end{abstract}
\end{titlepage}
\setcounter{page}{2}

\section{Introduction}
Two-particle correlations at low relative momenta (commonly referred to as \emph{femtoscopy}), which are sensitive to quantum statistics (in the case of identical particles) as well as strong and Coulomb final-state interactions (FSI), are used to extract the space-time characteristics of the particle-emitting sources created in
heavy-ion collisions~\cite{Goldhaber:1960sf,Kopylov:1972qw,Kopylov:1974uc}. The source radii extracted from these correlations describe the system at kinetic freeze-out, i.e.\ the last stage of particle interactions. Pion femtoscopy, which is the most common femtoscopic analysis, has shown
signatures of hydrodynamic flow in heavy-ion collisions, manifesting as a decrease in the source radii with increasing transverse mass
$m_{\rm T} = \sqrt{k_{\rm T}^2+m^2}$~\cite{Lisa:2005dd,Aamodt:2011mr}, where $k_{\rm T}=|\mathbf{p}_{\mathrm{T,1}}+\mathbf{p}_{\mathrm{T,2}}|/2$ is the average  transverse momentum of the pair. This behavior can be interpreted as one of the signatures of the formation
of deconfined quark matter in these collisions~\cite{Akkelin:1995gh}. However, a necessary condition for collective behavior
 is for ${\it all}$ particles created in the collision, not just pions, to experience
hydrodynamic flow. Thus, femtoscopic studies with particles other than pions are also needed.
It was shown that the hydrodynamic picture of nuclear collisions for the particular case of small transverse flow leads to the same $m_{\rm T}$ behavior of the longitudinal radii ($R_{\rm long}$) for pions and kaons~\cite{Makhlin:1987gm}. This common $m_{\rm T}$-scaling for $\pi$ and $\Kaon$ is an indication that the thermal freeze-out occurs simultaneously for $\pi$ and $\Kaon$ and that these two particle species are subject to the same Lorentz boost. Previous kaon femtoscopy studies carried out in Pb--Pb collisions at the SPS by the NA44, NA49, and CERES Collaborations~\cite{Bearden:2001sy, Afanasiev:2002fv,Adamova:2002wi} reported the decrease of $R_{\rm long}$ with $m_{\rm T}$ as $\sim m_{\rm T}^{-0.5}$ as a consequence of the boost-invariant longitudinal flow. Subsequent studies carried out in Au--Au collisions at RHIC~\cite{Adams:2003xp,Adamczyk:2013wqm,Afanasiev:2009ii} have shown the same level in the $m_{\rm T}$-dependencies for $\pi$ and $\Kaon$ radii, consistent with a common freeze-out hyper-surface. Like the SPS analysis, no exact universal $m_{\rm T}$-scaling for the 3D radii
was observed at RHIC. In the case of the one-dimensional correlation radius  $R_{\rm inv}$, only approximate scaling with $m_{\rm T}$ is expected as an additional confirmation of  hydrodynamic expansion~\cite{Lisa:2005dd}. In fact, $R_{\rm inv}$ source sizes as a function of $m_{\rm T}$ for different particle types ($\pi$, $\Kaon$, $\pro$...) follow the common curve with an accuracy of $\sim 10 \%$.

The motivation for comparing femtoscopic analyses with different particle species is not limited to studying $m_{\rm T}$ dependence.
The kaon analyses also offer a cleaner signal compared to pions, as they are less
affected by resonance decays, while the proton analysis provides a possibility for checking if baryons are included in the collective motion.
Studying charged and neutral kaon correlations together provides a convenient experimental consistency check, since they require
different detection techniques (charged tracks vs.\ decay vertex reconstruction) and call for different final-state interaction fitting parametrizations
(Coulomb-dominated vs. strong interaction-dominated), yet they are predicted to exhibit the same femtoscopic parameters~\cite{Shapoval:2014wya}. In addition to the charged kaon analyses at the SPS and RHIC, neutral kaon correlations were studied in Au-Au collisions at RHIC~\cite{Abelev:2006gu}, and ALICE has performed analyses on both charged and neutral kaons in $\pro\pro$ collisions~\cite{Abelev:2012ms,Abelev:2012sq}.
Recent pion femtoscopic results were obtained at RHIC~\cite{Adare:2015sca} and the LHC~\cite{Aamodt:2011mr,Abelev:2013pqa,Abelev:2014pja,CMS-PAS-HIN-14-013}, and proton femtoscopy has also been previously studied at RHIC~\cite{Gos:2007asd}.

This paper presents the results of femtoscopic studies of $\pi^{\pm}\pi^{\pm}$, $\Kpm\Kpm$, $\Kzs\Kzs$, $\pro \pro$, and $\pbar\pbar$ correlations from Pb-Pb collisions at
$\sqrt{s_{\mathrm {NN}}}=2.76$~TeV by the ALICE experiment at the LHC.
The femtoscopic radii and $\lambda$ parameters (the latter describe the decrease of the femtoscopic correlations due to e.g.\ long-lived resonances; see Sec.~\ref{sec:pions_femto} and~\ref{sec:results}) are extracted from
one-dimensional correlation functions in terms of the invariant momentum difference for a range of
collision centralities and $m_{\rm T}$ values. A hydrokinetic model~\cite{Shapoval:2014wya} is used to compare the kaon experimental results with hydrodynamic predictions.

The organization of the paper is as follows. In Sec.~\ref{sec:details}, we describe the data selection criteria. In Sec.~\ref{sec:fitting}, the details of the correlation
functions and the fitting process are discussed. The results of the analysis are shown in Sec.~\ref{sec:results}, and a summary is provided in Sec.~\ref{sec:summary}.

\section{Data analysis}
\label{sec:details}
The dataset analyzed in this paper is from Pb-Pb collisions at $\sqrt{s_{\rm NN}}=2.76$~TeV at the LHC measured by the ALICE detector~\cite{Aamodt:2008zz}.
About 8 million events from 2010 and about 40 million events from 2011 were used (2010 data were analyzed in the pion and $\Kzs$ analyses only). Events were classified according to their centrality determined using the measured amplitudes in the V0 detectors~\cite{Abelev:2013qoq}.
Charged particle tracking is generally performed using the Time Projection Chamber (TPC)~\cite{Dellacasa:2000bm} and the Inner Tracking System (ITS)~\cite{Aamodt:2008zz}. The ITS
allows for high spatial resolution in determining the primary (collision) vertex. In the pion, charged kaon, and proton analyses,
the determination of the momenta of the tracks was performed using tracks reconstructed with the TPC only and constrained to the primary vertex. Primary tracks were selected based on the distance of closest approach (DCA) to the primary vertex. Additional track selections based on the quality of the track reconstruction fit
and the number of detected ``hit'' points in the TPC were used. Also, all primary pairs sharing more than $5 \%$ of TPC clusters were rejected.
In the neutral kaon analysis, the secondary daughter tracks used global (TPC and ITS) track reconstruction and did not use any cuts based on track reconstruction quality or number of used or shared TPC clusters. The secondary vertex finder used to locate the neutral kaon decays employed the ``on-the-fly'' reconstruction method~\cite{Alessandro:2006yt}, which recalculates the daughter track momenta during the original tracking process under the assumption that the tracks came from a decay vertex instead of the primary vertex.

Particle identification (PID) for reconstructed tracks was carried out using both the TPC and the Time-of-Flight (TOF) detector~\cite{Cortese:2002kf} in the pseudorapidity range $|\eta|<0.8$. For TPC PID,
a parametrized Bethe-Bloch formula was used to calculate the specific
 energy loss (d$E$/d$x$)  in the detector expected for a particle with a given mass and momentum. For PID with TOF, the particle mass was used to calculate the expected time-of-flight as a function of track length and momentum.
For each PID method, a value $N_{\sigma}$ was assigned to each track denoting the number of standard deviations between the measured track information and the calculations
mentioned above. Different cut values of $N_{\sigma}$ were chosen based on detector performance for the various particle types and track momentum (see Table~\ref{tab:K0cuts} for
specific values used in each analysis)~\cite{Abelev:2014ffa}.

The analysis details specific to each particle species used in this study are discussed separately below.

\subsection{Pion selection}
The main single-particle selection criteria used in the pion analysis are summarized in Table~\ref{tab:K0cuts}.
Pion identification was performed using the TPC only.
An overall purity of the pion candidate sample was estimated using TPC d$E$/d$x$ distributions of the data and was found to be above 95\%. The main source of contamination comes from  $\rm e^{\pm}$ in the region where the d$E$/d$x$ curves for pions and electrons intersect.

Femtoscopic correlation functions of identical
particles are sensitive to the two-track reconstruction
efficiency because the correlated particle pairs (i.e.\ those with small relative momentum) generally have close trajectories. The main two-track issues are splitting (two tracks reconstructed from one particle) and merging (one track reconstructed from two particles), which are generally avoided using a track separation cut.
For pions, pairs were required to have a separation of $|\Delta \eta| > 0.016$ or \mbox{$\sqrt{\Delta\eta^2 +\Delta\varphi^{*2}} >0.045$} measured at the radial distance 1.2~m.
Here, $\eta$ is the pseudorapidity, and $\varphi^{*}$ is the
azimuthal coordinate taking into account track bending due to the magnetic field.

\subsection{Charged kaon selection}

\begin{figure}
  \begin{center}
    \includegraphics[width=.49\textwidth]{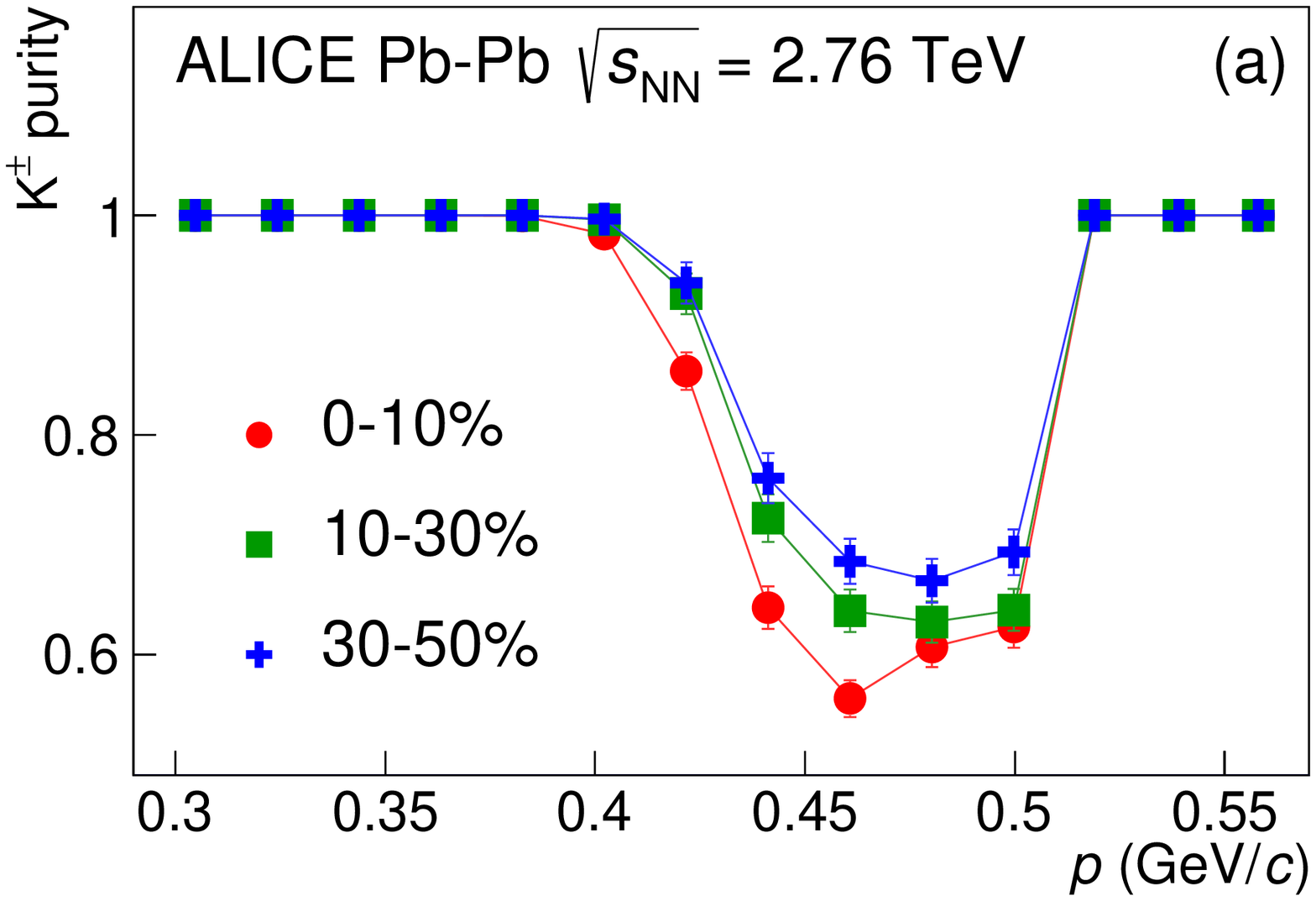}
    \includegraphics[width=.49\textwidth]{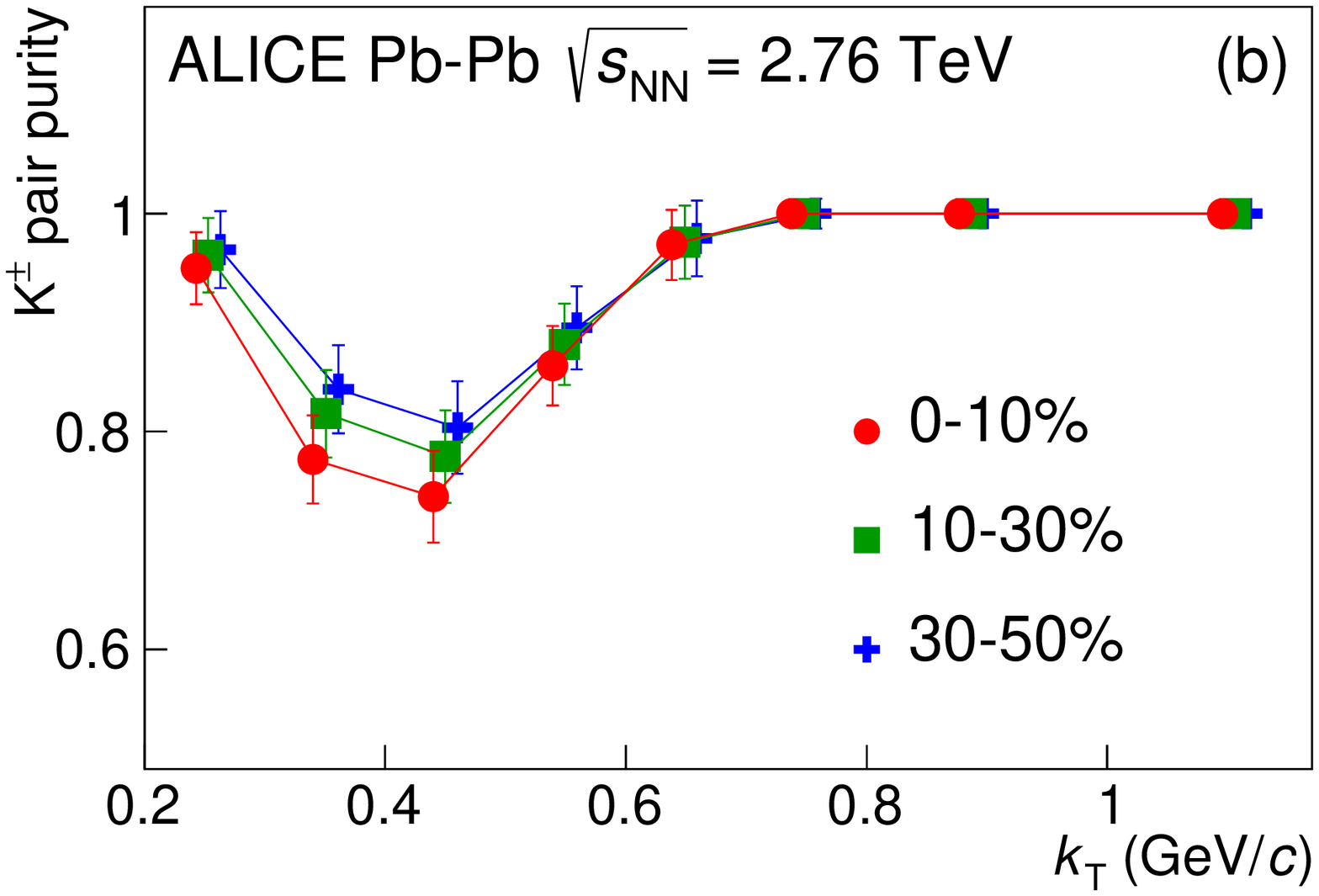}
    \caption{(Color online) Single $\Kpm$ purity (a) and $\Kpm$ pair
      purity (b) for different centralities. In (b) the $k_{\rm T}$ values for different centrality intervals are slightly offset for clarity.}
    \label{fig1}
  \end{center}
\end{figure}

 The main single-particle selection criteria used in the charged kaon analysis are listed in Table \ref{tab:K0cuts}. $\Kpm$ identification was performed using the TPC (for all momenta) and TOF (for $p > 0.5$ GeV/$c$) detectors.
Figure~\ref{fig1}~(a) shows the momentum dependence of the single kaon purity, defined as the fraction of accepted kaon tracks that correspond to true kaon particles. The purity values were obtained from TPC d$E$/d$x$ distributions of the data and by studying HIJING~\cite{Wang:1991hta} simulations using GEANT3~\cite{Brun:1994aa} to model particle transport through the detector.
Like the pions, the dominant contamination for charged kaons in the momentum region $0.4<p<0.5$~GeV/$c$ comes from $\rm e^{\pm}$.
The pair purity is calculated as a product of two single-particle purities,
where the momenta are taken from the experimental distribution.
The $\Kpm$ pair purity as a function of $k_{\rm T}$ for
three different centralities is shown in Fig.\ref{fig1}(b).

Regarding two-track selection criteria, charged kaon pairs were required to have a separation of \mbox{$|\Delta \eta|>0.02$} or \mbox{$|\Delta \varphi^{*}|>0.017$} measured at the radial distance 1.6~m.

\begin{table}
\centering
\begin{tabular}{l|l}
  \hline
    \multicolumn{2}{c}{Pion selection} \\ \hline
     Transverse momentum $p_{\rm T}$  & $0.14<p_{\rm T}<2.0$ GeV/$c$ \\ \hline
    $|\eta|$ & $< 0.8$ \\ \hline
    Transverse $\rm DCA$ to primary vertex & $< 0.2$ cm \\ \hline
    Longitudinal $\rm DCA$ to primary vertex & $< 0.15$ cm \\ \hline
    $N_{\sigma,\rm TPC}$ & $< 3$ \\ \hline
    \hline
  \multicolumn{2}{c}{Charged kaon selection} \\ \hline
    $p_{\rm T}$  & $0.15<p_{\rm T}<1.5$ GeV/$c$ \\ \hline
    $|\eta|$ & $< 0.8$ \\ \hline
    Transverse $\rm DCA$ to primary vertex & $< 2.4$ cm \\ \hline
    Longitudinal $\rm DCA$ to primary vertex & $< 3.0$ cm \\ \hline
    $N_{\sigma,\rm TPC}$ (for $p < 0.5$~GeV/$c$) & $< 2$ \\ \hline
    $N_{\sigma,\rm TPC}$ (for $p > 0.5$~GeV/$c$) & $< 3$ \\ \hline
    $N_{\sigma,\rm TOF}$ (for $0.5 < p < 0.8$ GeV/$c$) & $< 2$ \\ \hline
    $N_{\sigma,\rm TOF}$ (for $0.8 < p < 1.0$ GeV/$c$) & $< 1.5$ \\ \hline
    $N_{\sigma,\rm TOF}$ (for $1.0 < p < 1.5$ GeV/$c$) & $< 1.0$ \\ \hline
  \hline
  \multicolumn{2}{c}{Neutral kaon selection} \\ \hline
    $|\eta|$ & $< 0.8$ \\ \hline
    Daughter-daughter $\rm DCA_{3D}$ & $< 0.3$ cm \\ \hline
    $\rm DCA_{3D}$ to primary vertex & $< 0.3$ cm \\ \hline
    Decay length & $< 30$ cm \\ \hline
    Cosine of pointing angle & $> 0.99$ \\ \hline
    Invariant mass & $0.480 < m_{\pi^+ \pi^-} < 0.515$ GeV/${c}^2$ \\ \hline
    Daughter $p_{\rm T}$ & $> 0.15$ GeV/c \\ \hline
    Daughter $|\eta|$ & $< 0.8$ \\ \hline
    Daughter $\rm DCA_{3D}$ to primary vertex & $> 0.4$ cm \\ \hline
    Daughter $N_{\sigma,\rm TPC}$ & $< 3$ \\ \hline
    Daughter $N_{\sigma,\rm TOF}$ (for $p > 0.8$ GeV/$c$) & $< 3$ \\ \hline
  \hline
  \multicolumn{2}{c}{Proton selection} \\ \hline
    $p_{\rm T}$  & $0.7<p_{\rm T}<4.0$ GeV/$c$ \\ \hline
    $|\eta|$ & $< 0.8$ \\ \hline
    Transverse $\rm DCA$ to primary vertex & $< 2.4$ cm \\ \hline
    Longitudinal $\rm DCA$ to primary vertex & $< 3.2$ cm \\ \hline
    $N_{\sigma,\rm TPC}$ (for $p < 0.8$~GeV/$c$) & $< 3$ \\ \hline
    $\sqrt{{N_{\sigma,\rm TPC}^2+N_{\sigma,\rm TOF}^2}}$ (for $p > 0.8$~GeV/$c$) & $< 3$ \\ \hline
\end{tabular}
\caption{Single particle selection criteria.}
\label{tab:K0cuts}
\end{table}

\subsection{Neutral kaon selection}
\begin{figure}
\begin{center}
\includegraphics[width=.90\textwidth]{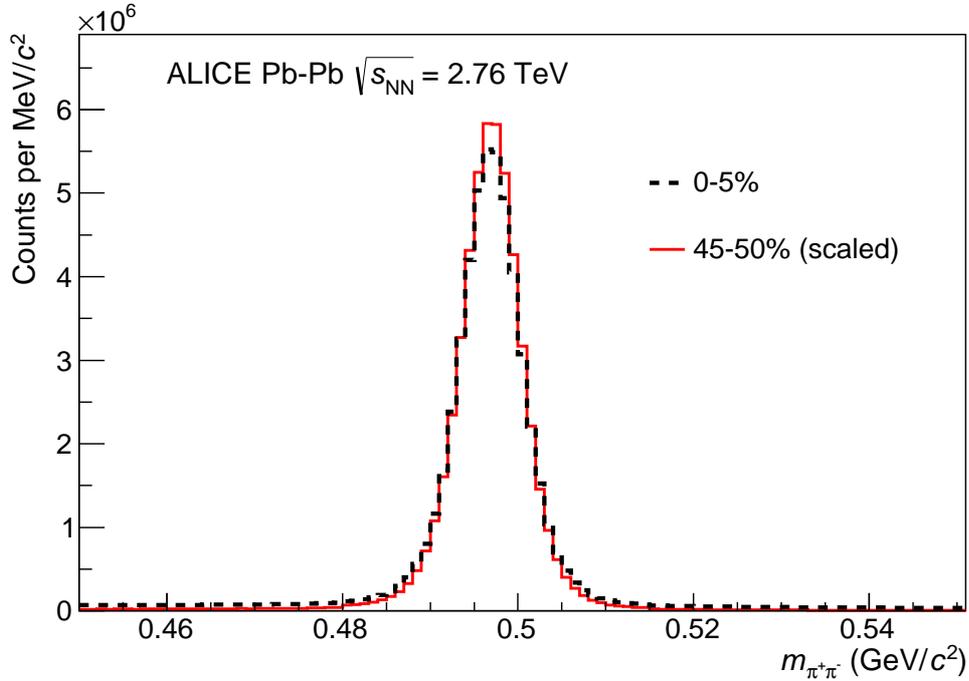}
\caption{(Color online) Invariant mass distribution of $\pi^+\pi^-$ pairs showing the $\Kzs$ peak for two centrality intervals.
The 45-50\% centrality is scaled so that both distributions have
the same integral in the range \mbox{$0.480 < m_{\pi^+ \pi^-} < 0.515$}~GeV/${\rm c}^2$.}
\label{fig2}
\end{center}
\end{figure}

The decay channel $\Kzs\rightarrow\pi^+\pi^-$ was used for the identification of neutral kaons. The single-particle cuts for parents ($\Kzs$) and daughters ($\pi^{\pm}$) used in the decay-vertex reconstruction are shown in Table \ref{tab:K0cuts}.
PID for the pion daughters was performed using both TPC (for all momenta) and TOF (for $p>0.8$~GeV/$c$). Figure \ref{fig2} shows an example of the $\pi^+\pi^-$ invariant mass distribution where the
$\Kzs$ peak is seen. The cuts used in this analysis were chosen to balance statistics and signal purity. The neutral kaon purity (defined as Sig./[Sig.+Bkg.]\ for $0.480 < m_{\pi^+ \pi^-} < 0.515$~GeV/${\rm c}^2$) was found to be greater than 0.95.

Two main two-particle cuts were used in the neutral kaon analysis. To resolve two-track inefficiencies associated with the daughter tracks, such as the splitting or merging of tracks discussed above,
a separation cut was employed in the following way. For each kaon pair, the spatial separation between the same-sign pion daughters
was tabulated at several points throughout the TPC (every 20 cm radially from 85 cm to 245 cm) and averaged.
If the average separation of either pair of tracks was below 5 cm, the kaon pair was not used. Another cut was used to prevent two reconstructed kaons from using the same daughter track. If two kaons shared a daughter track,
one of them was cut using a procedure which compared the two $\Kzs$ candidates and kept the candidate
whose reconstructed parameters best matched those expected of a true $\Kzs$ particle in two of three categories
(smaller $\Kzs$ DCA to primary vertex, smaller daughter-daughter DCA, and $\Kzs$ mass closer to the PDG value~\cite{Agashe:2014kda}).
This procedure was shown, using HIJING+GEANT3 simulations, to have a success rate of about 95\% in selecting a true $\Kzs$ particle over a fake one.
More details about $\Kzs\Kzs$ analysis can be found in Ref.~\cite{Abelev:2012ms,MattThesis}.

\subsection{Proton selection}
 The single-particle cuts used in the
proton analysis are summarized in Table~\ref{tab:K0cuts}.
The proton analysis used tracks with $0.7 < p_{\mathrm{T}} < 4.0$~GeV/$c$
. The lower $p_{\mathrm{T}}$ cut is used to suppress protons coming
from weak decays and interactions with the detector material.
Particle identification for $\pro$ and $\pbar$ was performed using both TPC (for all momenta) and TOF (for $p>0.8$~GeV/$c$).
The proton purity was estimated
using HIJING+GEANT3 simulations and was found to be greater than $95\%$. The used DCA criteria do not fully discriminate between primary protons and protons from weak decays. This may lead to a significant contamination from protons from lambda particles. The effect of this contamination is discussed in Sec.~\ref{sec:protons_femto}.

Regarding two-track selection criteria, pairs were required to have a separation of $|\Delta \eta| >0.01$ or $|\Delta \varphi^{*}|>0.045$ measured at the radial distance 1.2~m.

\section{Construction of the correlation functions and fitting procedures}
\label{sec:fitting}
The experimental two-particle correlation function is defined as $C(\textbf{q})=A(\textbf{q})/B(\textbf{q})$, where $A(\textbf{q})$ is the measured distribution
of same-event pair momentum difference, $\textbf{q}=\textbf{p}_1-\textbf{p}_2$, and $B(\textbf{q})$ is the reference distribution of pairs from mixed events.
The pairs in the denominator distribution $B(\textbf{q})$ are constructed by taking a particle from one event and pairing it with a particle from another
event with a similar centrality and primary vertex position along the beam direction. Each event is mixed with five (ten) others for the $\Kzs$ ($\pi^{\pm}$, $\Kpm$, $\pro$) analysis.
The available statistics of proton pairs with low $q$ ($< 0.2$~GeV/$c$) allowed us to perform the analyses only for the one-dimensional correlation
function $C(q)$, where $q = \left|\textbf{q}\right|$ in the Pair Rest Frame (PRF). In the case of pions and kaons, the statistics were high enough for three-dimensional studies, but these are beyond the scope of this paper; here, only the one-dimensional analysis is presented in order to compare results with heavier particles. The numerator and denominator are normalized such that $C(q) \rightarrow 1$ as $q \rightarrow \infty$. Pair cuts have been applied in exactly the same
way for the same-event (signal) and mixed-event (background) pairs.

All correlation functions have been corrected for momentum resolution effects. The correction factors
were determined using HIJING events to build simulated correlation functions using theoretical correlation functions as weights. The ratio of the correlation functions using HIJING generated momenta to those using HIJING+GEANT3 reconstructed momenta forms the correction factor.

For the analyses presented in this paper, the theoretical femtoscopic correlation function is defined as the square of the two-particle wavefunction averaged over the relative
distance $\textbf{r}^*$ of the emitters in the PRF. This is performed using the Koonin-Pratt equation~\cite{Koonin:1977fh,Pratt:1990zq}
\begin{equation}
C(q) = \int S(\mathbf{r}^*) \lvert \Psi (q,\mathbf{r}^*) \rvert ^2 {\rm d^3} \mathbf{r}^*.
\label{eq:KooninPratt}
\end{equation}

For the one-dimensional analysis, we assume a spherically symmetric Gaussian distribution of the particle emitter spatial separation ${\bf r^*}$ in the PRF with size $R_{\rm inv}$~\cite{Lednicky:1981su},
\begin{equation}
\label{eq:space}
 S(\mathbf{r}^*)\sim \exp(-{\bf r}^{*2}/4R^2_{\rm inv}).
\end{equation}
The two-particle wavefunction is (anti-)symmetrized for identical bosons (fermions) and may include terms incorporating Coulomb or strong
final-state interactions, depending on the type of particles being studied.

The methods used in constructing and fitting the various correlation functions
are discussed separately below.

\subsection{Pions}
\label{sec:pions_femto}
Pion correlation functions were fitted using the Bowler-Sinyukov formula~\cite{Bowler:1991vx,Sinyukov:1998fc}:
\begin{equation}
C(q)= N \left[1 -\lambda +\lambda K(q)\left( 1+
\exp{\left(-R_{\rm inv}^{2} q^{2}\right)}\right)\right],
\label{eq:BS}
\end{equation}
where $N$ is the normalization factor. The $\lambda$ parameter (also used in the other analyses) can be affected by long-lived resonances, coherent sources~\cite{PhysRevC.65.064904,Abelev:2013pqa,Wiedemann:1996ig}, and non-Gaussian features of the particle-emission distribution.
$K(q)$ is a symmetrized $K$-factor calculated according to Ref.~\cite{Sinyukov:1998fc,Abelev:2013pqa} as
\begin{equation}
K(q) = C({\rm QS+ Coulomb}) / C(\rm QS), \label{eq:newCoulomb}
\end{equation}
where $C(\rm QS)$ and $C(\rm QS+\rm Coulomb)$ are the theoretical correlation
functions calculated with THERMINATOR~2~\cite{Chojnacki:2011hb} using the quantum statistics
(``QS'') and ``QS+Coulomb'' weights (i.e.\ squared wavefunction), respectively~\cite{Lednicky:2005af}.
The effect of the strong interaction is neglected here, since for like-sign pions, the contribution is  small for the expected source sizes~\cite{Lednicky:2005af}. Figure \ref{figAdded} shows an example $\pi^+\pi^+$ correlation function with the corresponding line of best fit. More details about the pion analysis may be found in~\cite{Adam:2015vna}.

\subsection{Charged kaons}
\label{sec:charged_kaon_femto}
\begin{figure}
\begin{center}
\includegraphics[width=.9\textwidth]{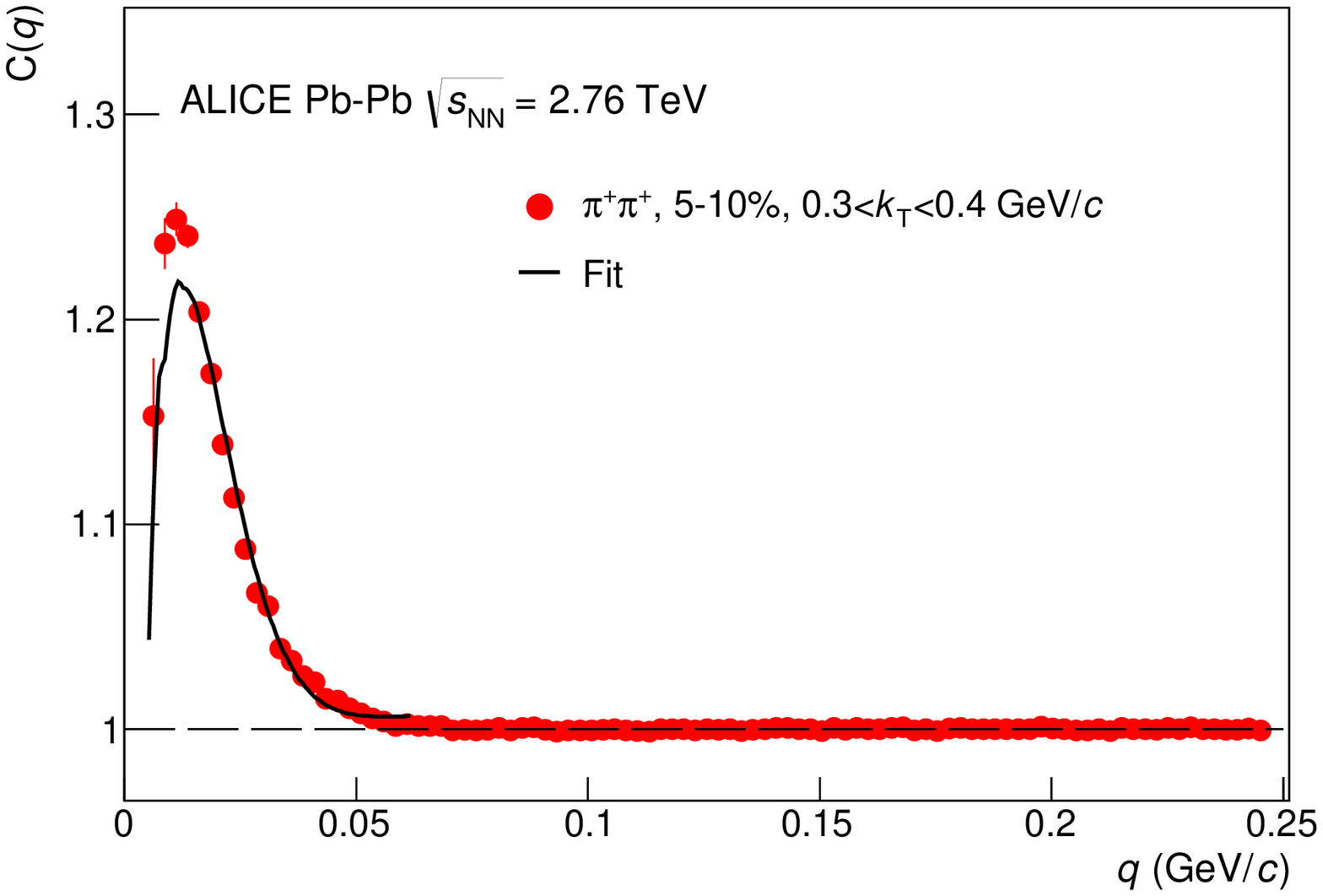}
\caption{(Color online) Example correlation function with fit for
$\pi^+\pi^+$ for centrality 5-10\% and $\left < k_{\rm T} \right > = 0.35$~GeV/$c$. Statistical uncertainties are shown as thin lines.}
\label{figAdded}
\end{center}
\begin{center}
\includegraphics[width=.9\textwidth]{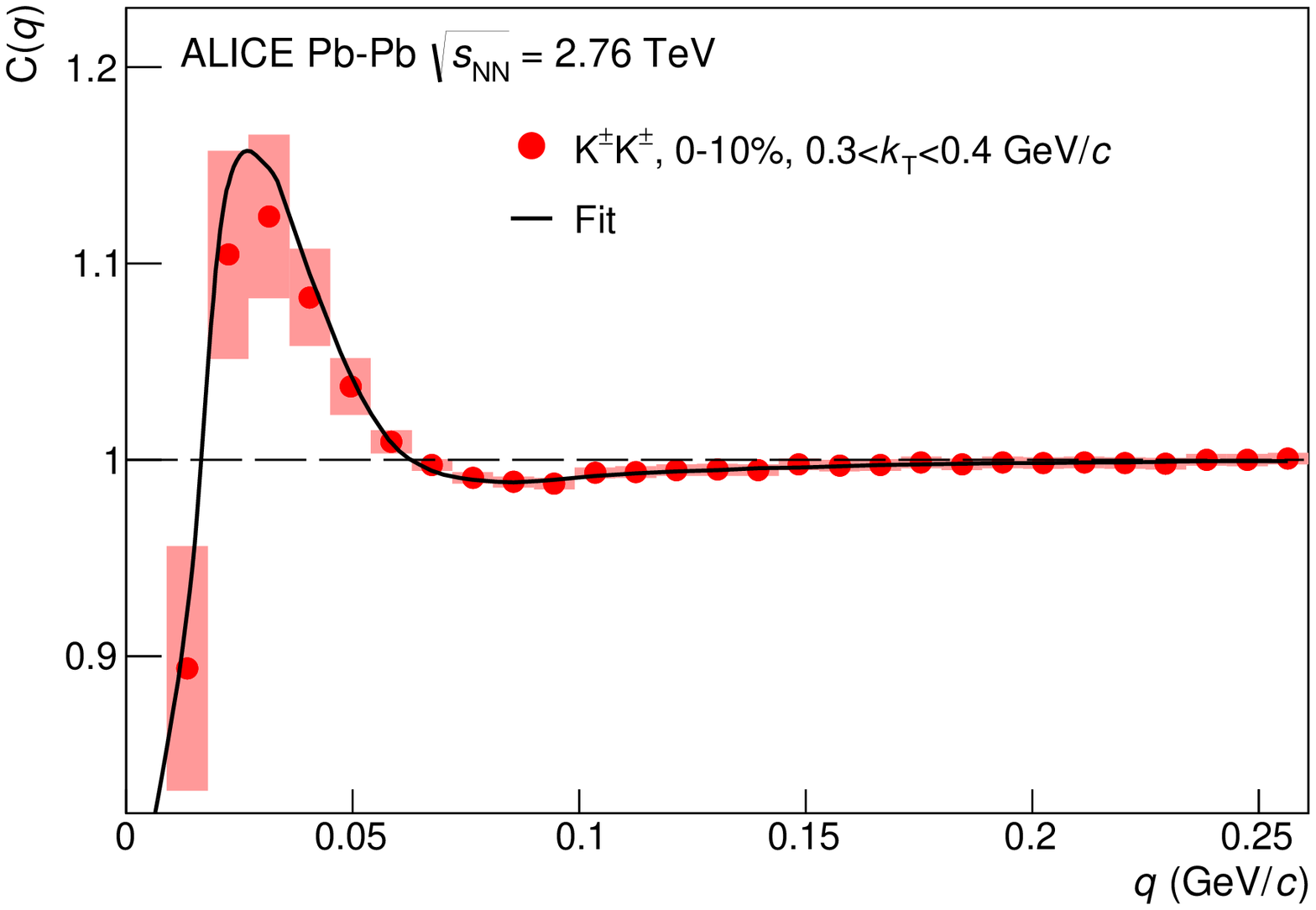}
\caption{(Color online) Example correlation function with fit for
$\Kpm\Kpm$ for centrality 0-10\% and $\left < k_{\rm T} \right > = 0.35$~GeV/$c$. Systematic uncertainties (boxes) are shown; statistical uncertainties are within the data markers. The main sources of systematic uncertainty are the momentum resolution correction and PID selection.}
\label{fig4a}
\end{center}
\end{figure}

Figure \ref{fig4a} shows an example $\Kpm\Kpm$ correlation function with the corresponding line of best fit.
A purity correction was applied to the correlation function according to
\begin{equation}
C_{\rm corrected} = (C_{\rm raw}
-1+P)/P,
\label{eq:puritycorrection}
\end{equation}
where the pair purity $P$ is taken from Fig.~\ref{fig1}.
$\Kpm\Kpm$ correlation functions were fitted using the Bowler-Sinyukov formula of Eq.~\ref{eq:BS}; the procedure is essentially the same as for pions. There are no available experimental data for $\Kpm\Kpm$ strong FSI. The influence of the strong interaction to the correlation function was estimated with the s-wave scattering length calculated within the fully-dynamical lattice QCD~\cite{Beane:2007uh}. The systematic uncertainty assigned to this effect was determined to be 4\%.

\begin{figure}
\begin{center}
\includegraphics[width=.9\textwidth]{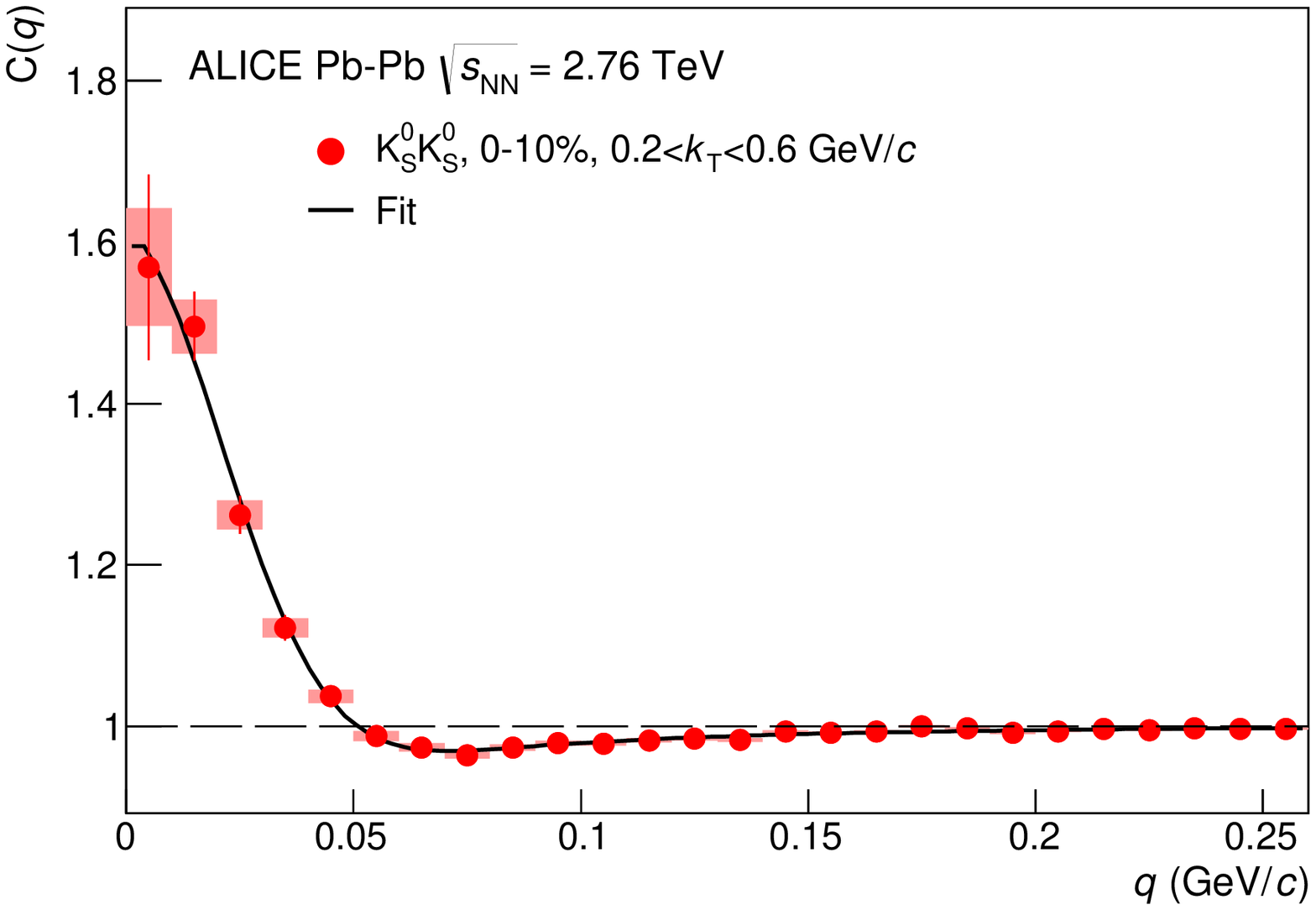}
\caption{(Color online) Example correlation function with fit for
$\Kzs\Kzs$ for centrality 0-10\% and $\left < k_{\rm T} \right > = 0.48$~GeV/$c$.  Statistical (thin lines) and systematic (boxes) uncertainties are shown. The main source of systematic uncertainty is the variation of single-particle cuts.}
\label{fig4b}
\end{center}
\begin{center}
\includegraphics[width=.9\textwidth]{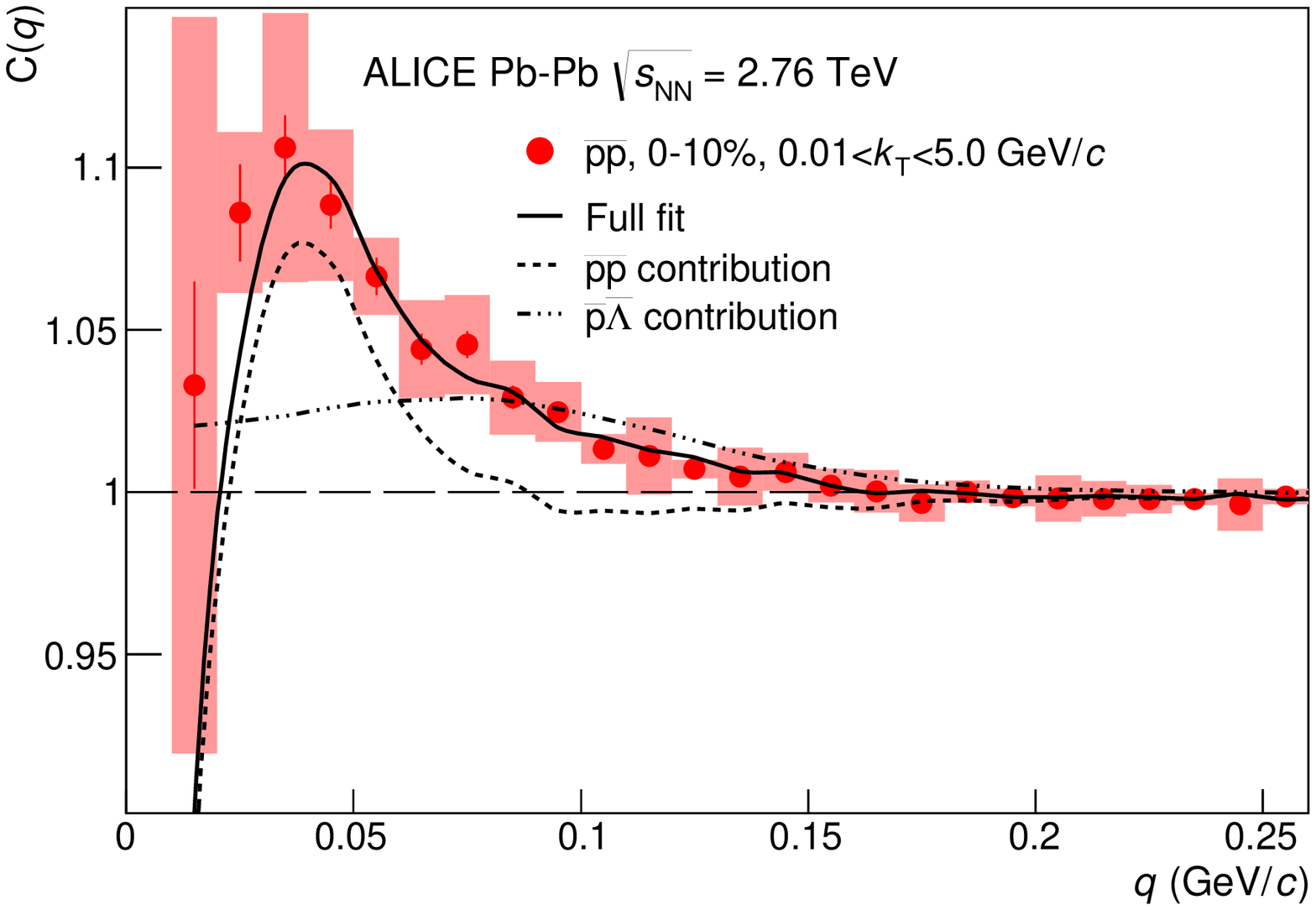}
\caption{(Color online) Example correlation function with fit for
$\pbar\pbar$ for centrality 0-10\% and $\left < k_{\rm T} \right > = 1.0$~GeV/$c$. Statistical (thin lines) and systematic (boxes) uncertainties are shown. The main source of systematic uncertainty is the variation of two-track cuts.  }
\label{fig4c}
\end{center}
\end{figure}

\subsection{Neutral kaons}
Figure~\ref{fig4b} shows an example $\Kzs\Kzs$ correlation function with the corresponding line of best fit.
$\Kzs\Kzs$ correlation functions were fitted with a parametrization which includes Bose-Einstein statistics as well as strong final-state interactions (FSI)~\cite{Lednicky:1981su,Abelev:2006gu},

\begin{equation}
C(q)=\left[1-\lambda+\lambda C'(q)\right]\left(a+bq\right),
\label{eq:fiteqKs1}
\end{equation}
where
\begin{equation}
C'(q)=1+e^{-q^2R^2}+C_{\rm strong FSI}(q,R),
\label{eq:fiteqKs2}
\end{equation}
\begin{equation}
 C_{\rm strong FSI}(q,R) =  \dfrac{1}{2} \left [ \left| {f(q)}\over{R}\right|^2
+ {{4 \Re f(q)}\over{\sqrt{\pi} R}} F_1 (q R) - {{2\Im f(q)}\over{R}}F_2(q R) \right ],
\label{eq:strongfsi}
\end{equation}
and
\begin{equation}
F_1(z)=\int^{z}_{0}{\rm d}x\frac{e^{x^2-z^2}}{z}; \;\; F_2(z)=\frac{1-e^{-z^2}}{z}.
\label{eq:fiteqKs3}
\end{equation}
$f(q)$ is the s-wave scattering amplitude for the $\Kz\Kzb$ system; we neglect the scattering for $\Kz\Kz$ and $\Kzb\Kzb$ due to small scattering lengths $\approx 0.1$~fm~\cite{Abelev:2006gu}. The factor of 1/2 in Eq.~\ref{eq:strongfsi} is due to the fact that half of the $\Kzs\Kzs$ pairs come from $\Kz\Kzb$.
The strong FSI have a significant effect on the $\Kz\Kzb$ contribution to the $\Kzs\Kzs$ correlation function due to the near-threshold resonances,
$\fz$(980) and $\az$(980). For the scattering amplitude, only s-wave contributions were taken into account; the higher-order corrections were small and therefore neglected~\cite{Lednicky:2005tb}. The scattering amplitude $f(q)$ is calculated using a a two-channel parametrization which accounts for the elastic transition
$\Kz \Kzb \rightarrow \Kz \Kzb$ and the inelastic transition $\Kplus \Kmin \rightarrow \Kz \Kzb$ (See Ref.~\cite{Abelev:2006gu} for more
detailed expressions describing the fit function). Equation~\ref{eq:fiteqKs1} also includes an additional factor to account for non-femtoscopic background correlations at large $q$,
with $a$ and $b$ being free parameters in the fit.

\subsection{Protons}
\label{sec:protons_femto}
Figure~\ref{fig4c} shows an example $\pbar\pbar$ correlation function with the corresponding line of best fit.
The femtoscopic correlations of $\pro\pro$ and $\pbar\pbar$ pairs are due to a combination of Fermi-Dirac statistics, Coulomb and strong FSI.
A distinct maximum is seen at ${ q} \approx 40$~MeV/$c$~\cite{Lednicky:1981su}; this enhancement is due to the strong interaction, as both quantum statistics and Coulomb interaction present a negative correlation.
Due to the fact that feed-down from weak decays cannot be neglected in high-energy heavy-ion collisions, the effects of residual correlations related
to the $\pro\Lambda$ system are taken into account. The proton daughter of a $\Lambda$ decay has similar momentum to the $\Lambda$ itself and may survive the experimental selection for primary protons. Thus, it may contribute to the measured correlations by forming a pair with a primary proton. As can be seen in Fig.~\ref{fig4c}, attempting to fit the measured correlation functions with the theoretical $\pro\pro$ ($\pbar\pbar$) functions alone was unsuccessful due to the additional positive correlation
observed in the range \mbox{$60 < q < 160$~MeV/$c$}. Thus, a method of simultaneous fitting of
$\pro\pro$ ($\pbar\pbar$) and $\pro\Lambda$ ($\pbar\overline{\Lambda}$) correlations was applied.
Contributions from heavier baryon-baryon pairs are not taken into account since the original correlation between the parent particles is not known due to unknown interaction parameters, for example for the $\Lambda\Lambda$ pair. Moreover such residual correlations are more smeared compared with $\pro\Lambda$ because of larger decay momentum. In addition, the fraction of baryons heavier than $\Lambda$ decaying to protons is smaller than the fraction of $\Lambda$'s. Finally, comparing with baryon-antibaryon pairs analysed in~\cite{Kisiel:2014mma}, the width of the correlation for baryon-baryon pairs is much smaller, and therefore the effect is much more smeared due to decay kinematics.

The experimental correlation function of $\pro\pro$ and $\pbar\pbar$ systems were fitted with~\cite{Kisiel:2014mma}
\begin{equation}
  C_{\rm meas} (q_{\pro\pro}) = 1 + \lambda_{\pro\pro}  (C_{\pro\pro} (q_{\pro\pro};R) - 1) + \lambda_{\pro\Lambda} (C_{\pro\Lambda} (q_{\pro\pro};R) - 1),
  \label{fitformula}
\end{equation}
where
$\lambda_{\pro\pro}$ is the fraction of correlated $\pro\pro$ pairs where both particles are primary, and
$\lambda_{\pro\Lambda}$ is the fraction of correlated $\pro\pro$ pairs where one particle is primary and the other is a daughter of $\Lambda$ decay.
The theoretical proton-proton correlation function 
was calculated as
\begin{equation}
C(q_{\pro\pro}) = {1 \over 4} \left[ {{\int S({\mathbf{r^{*}}})
  {1 \over 2}|\Psi^{\rm S}_{-\mathbf{q}_{\pro\pro}}({\mathbf{r^{*}}})+\Psi^{\rm S}_{+\mathbf{q}_{\pro\pro}}({\mathbf{r^{*}}})|^{2}} \over {\int
  S({\mathbf{r^{*}}}})} \right]
+{3 \over 4} \left[  {{\int S({\mathbf{r^{*}}})
  {1 \over 2}|\Psi^{\rm T}_{-\mathbf{q}_{\pro\pro}}({\mathbf{r^{*}}})-\Psi^{\rm T}_{+\mathbf{q}_{\pro\pro}}({\mathbf{r^{*}}})|^{2}} \over {\int
  S({\mathbf{r^{*}}}})}\right].
\label{eq:cfrompsi}
\end{equation}
This formulation takes into account the necessary (anti-)symmetrization of the wavefunction for a $\pro\pro$ pair in the singlet (triplet) spin state with a corresponding weight of 1/4 (3/4).
The $\pro\pro$ pair wavefunction may be written as~\cite{Lednicky:2005tb}
 \begin{equation}
 \Psi_{-\mathbf{q}_{\pro\pro}}({ \mathbf{r^{*}}}) = \mathrm{e}^{\mathrm{i} \delta_c} \sqrt{A_{\rm c}(\eta)} \left [ \mathrm{e}^{-\mathrm{i} \mathbf{q}_{\pro\pro}\cdot\mathbf{r^{*}}/2} F(- \mathrm{i}\eta,1,\mathrm{i} \xi)  + f_{\rm c}(q_{\pro\pro}) {{\tilde{G}(\rho,\eta)}\over{\lvert\mathbf{r^{*}}\rvert}} \right ],
 \label{eq:psidef}
 \end{equation}
where $\mathbf{r^{*}}$ is the spatial separation of particle emitters at generally different emission moments in the PRF, $\delta_c=\mathrm{arg}\, \Gamma(1+\mathrm{i}\eta)$ is the Coulomb s-wave phase shift, $A_{\rm{c}}{(\eta)} = 2 \pi \eta (\mathrm{e}^{2 \pi \eta}-1)^{-1}$ is the Gamow factor (also referred to as Coulomb penetration factor), $\eta = (\frac{1}{2}aq_{\pro\pro})^{-1}$, $a=(\mu z_1 z_2 e^2)^{-1}$ is the two-particle Bohr radius taking into account the sign of the interaction ($a=57.6$~fm for $\pro\pro$ pair), $F$ is the confluent hypergeometric function, $\xi={1\over2}q_{\pro\pro}r^{*}(1+\cos{\theta^{*}})$, $\theta^{*}$ is the angle between $\mathbf{q}_{\pro\pro}$ and $\mathbf{r^{*}}$, $\tilde{G}$ is the combination of the regular and singular s-wave Coulomb functions,  and $\rho={1\over2}q_{\pro\pro}r^{*}$. The amplitude of the low-energy s-wave elastic scattering due to the short range interaction $f_c(q_{\pro\pro})$ may be expressed as
 \begin{equation}
f_c(q_{\pro\pro})=\left [ {1 \over f_0} + {d_0 q_{\pro\pro}^2 \over 8} - {1 \over 2} \mathrm{i} q_{\pro\pro} A_{\rm c} (\eta) - {2 \over a} h(\eta)\right ]^{-1},
 \label{eq:fdef}
 \end{equation}
where $f_0$ is the scattering length, $d_0$ is the effective radius of the interaction, $h(\eta) = [\psi(\mathrm{i}\eta)+\psi(-\mathrm{i}\eta)-\mathrm{ln}(\eta^2)]/2$, and $\psi$ is the digamma function. For the $\pro\pro$ system in the singlet (triplet) state, $f_0$ and $d_0$ are 7.77~fm (-5.4~fm) and 2.77~fm (1.7~fm).

For the feed-down term, the theoretical $\pro\Lambda$ correlation function for a given $R_{\pro\Lambda}$ transformed into the $\pro\pro$ momentum space is obtained from the Lednicky-Lyuboshitz model~\cite{Lednicky:1981su} and calculated as
\begin{equation}
C_{\pro\Lambda} (q_{\pro\pro};R_{\pro\Lambda}) = \sum\limits_{q_{\pro\Lambda}} {C_{\pro\Lambda} (q_{\pro\Lambda};R_{\pro\Lambda})
T(q_{\pro\pro},q_{\pro\Lambda})}/\sum\limits_{q_{\pro\Lambda}} {T(q_{\pro\pro},q_{\pro\Lambda})},
\end{equation}
where \mbox{$C_{\pro\Lambda}(q_{\pro\Lambda};R_{\pro\Lambda}) = 1+C_{\rm strong FSI}(q_{\pro\Lambda};R_{\pro\Lambda})$}, and $T(q_{\pro\pro},q_{\pro\Lambda})$ are the transformation factors
related to $\Lambda$ decay kinematics, calculated with THERMINATOR~2~\cite{Chojnacki:2011hb}. Here, a spin-dependent version of Eq.~\ref{eq:strongfsi} is used~\cite{Lednicky:1981su}:
\begin{equation}
C_{\rm strong FSI}(q,R) = \sum_{S} \rho_S  \left [{1 \over 2} \left| {f^S(q)}\over{R}\right|^2  \left( 1-{{d_0^S}\over{2 \sqrt{\pi} R} } \right)
+ {{2 \Re f^S(q)}\over{\sqrt{\pi} R}} F_1 (q R) - {{\Im f^S(q)}\over{R}}F_2(q R) \right ],
\end{equation}
where $f^S(q)$ is the spin-dependent scattering amplitude, $\rho_s$ is the fraction of pairs in each total spin state $S$, and $d_0^S$ is the effective radius of the interaction.
It is assumed that the radii of $\pro\pro$ 
and $\pro\Lambda$ 
sources are equal. Therefore, there are three free fit parameters in Eq.~\ref{fitformula}: $\lambda_{\pro\pro}$, $\lambda_{\pro\Lambda}$ 
, and $R$. Theoretical $\pro\pro$ and $\pro\Lambda$ correlation functions were calculated using several values of the free parameters,
 and the fit function (for the set of parameters given during each fit iteration) was formed by a quadratic interpolation
of the calculated correlation functions.

\subsection{Systematic uncertainties}

The effects of various sources of systematic uncertainty on the extracted fit parameters were studied as functions of centrality and $k_{\rm T}$. Table~\ref{tab:systerr} shows the minimum and maximum uncertainties from each source. The values of the total uncertainty are not necessarily equal to the sum of the  individual uncertainties, as the latter can come from different centrality or $k_{\rm T}$ bins.
All four analyses studied the effects of changing the selection criteria for the events, particles and pairs used (variation of cut values up to $\pm$25$\%$) and varying the range of $q$ values over which the fit is performed (variation of $q$ limits up to $\pm$25$\%$). Uncertainties associated with momentum resolution corrections are included
in the $\pi$, $\Kpm$, and $\pro$ analyses; the $\Kzs$ analysis also studied this and found the uncertainties to be negligible. The $\Kpm$, $\Kzs$, and $\pro$ analyses encountered uncertainties associated with the non-flat background seen at large-$q$ for high-$k_{\rm T}$ pairs in peripheral collisions (estimated by using different parametrizations (linear or polynomial) to fit the large-$q$ region). Strong FSI uncertainties affect both kaon analyses. For $\Kzs$, the strong FSI uncertainty comes from the fact that several sets of $\rm f_0(980)$ and $\rm a_0(980)$ parameters are available~\cite{Martin1977,Antonelli2002,Achasov1,Achasov2}; each set is used to fit the data, the results are averaged, and the maximum difference was taken as the systematic error. The $\pi$ and $\Kpm$ analyses have uncertainties associated with the choice of the Coulomb function used in the fitting procedure. The $\Kpm$ analysis had additional uncertainties due to the misidentification of particles and the associated purity correction.
The $\pro$ analysis also had uncertainties associated with the uncertainty in the $R_{\pro\pro}/R_{\pro\Lambda}$ ratio and attempts to fix
the $\lambda_{\pro\pro}$ parameter using the single-particle purity. All of the analyses were performed separately for the two different signs of the ALICE dipole magnetic field, but the resulting systematic uncertainty was found to be negligible in all cases.

Systematic uncertainties on correlation functions~(Figs.~\ref{fig4a},~\ref{fig4b} and~\ref{fig4c}) were derived from the variation of single- and two-particle cuts.
\begin{table}
 \centering
 \begin{tabular}{cc|c|c|c|c|c|c|c}
   & \multicolumn{2}{c}{$\pi^{\pm}$} & \multicolumn{2}{c}{$\Kpm$} & \multicolumn{2}{c}{$\Kzs$} & \multicolumn{2}{c}{$\pro$} \\ \cline{2-9}
   \multicolumn{1}{c}{} & $R_{\rm inv}$ & $\lambda$& $R_{\rm inv}$ & $\lambda$ & $R_{\rm inv}$ & $\lambda$ & $R_{\rm inv}$ & $\lambda$ \\ \hline
   \multicolumn{1}{l}{Event/particle/pair selection} & 2-13 & 6& 3-5 & 3-5 & 1-4 & 2-10 & 2-27
                                                                   & 12-58 \\ \hline
   \multicolumn{1}{l}{Non-flat background} & - & -& 0.2-5 & 0.2-5 & 0-5 & 0-4 & 2-3
                                                                   & 1-9 \\ \hline
   \multicolumn{1}{l}{Fit range} & 10 & 33& 1-5 & 1-5 & 0-4 & 0-3 & 3-26 
                                                                   & 3-57 \\ \hline
   \multicolumn{1}{l}{Strong FSI} & - & -& 4 & 4 & 1-2 & 5-10 & n/a & n/a \\ \hline
   \multicolumn{1}{l}{Coulomb function} & 3 & 3& 2 & 4 & n/a & n/a & n/a & n/a \\ \hline
   \multicolumn{1}{l}{PID and purity} & - & -& - & 5-28 & - & - & 4-18 & 13-41  \\ \hline
   \multicolumn{1}{l}{Momentum resolution} & 2 & 3& 3-5 & 5-10 & - & - & 3 & 1-8 \\ \hline
   \multicolumn{1}{l}{Fixing $\lambda_{\pro\pro}$} & n/a & n/a& n/a & n/a & n/a & n/a & 1-29
                                                                   & n/a \\ \hline
   \multicolumn{1}{l}{$R_{\pro\pro}/R_{\pro\Lambda}$ ratio} & n/a & n/a& n/a & n/a & n/a & n/a & 1-13
                                                                   & 20-52 \\ \hline
   \multicolumn{1}{l}{\bf Total (quad. sum)} & 11-21 & 34& 6-9 & 10-32 & 2-7 & 7-15 & 10-40 & 30-80 \\ \hline
 \end{tabular}
 \caption{Minimal and maximal uncertainty values for various sources of systematic uncertainty (in percent). The $\lambda$ for the proton analysis refers to the sum
   of $\lambda_{\pro\pro}$ and $\lambda_{\pro\Lambda}$. Please note that each value is the minimum (maximum) uncertainty from a specific source, but each can be from a different centrality or $k_{\rm T}$ bin. Thus, the minimum (maximum) total uncertainties are greater (smaller) than (or equal to) the sum of the minimum (maximum) individual uncertainties. 'n/a' denotes that the given descriptor of the systematic uncertainty is not applicable for the specific pair type, and '-' means that the contribution from the given source is negligible.
 }
 \label{tab:systerr}
\end{table}

\section{Results}
\label{sec:results}
\begin{figure}
\begin{center}
\includegraphics[width=0.90\textwidth]{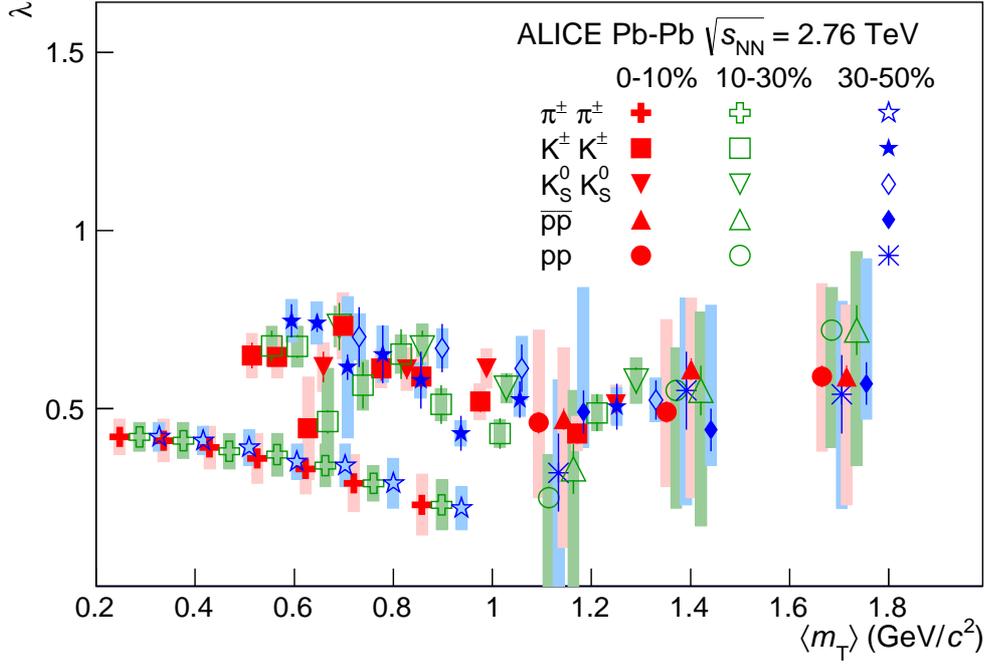}
\caption{(Color online) $\lambda$ parameters ($\lambda_{\pro\pro}+\lambda_{\pro\Lambda}$ in case of (anti)proton pairs) vs.\ $m_{\rm T}$ for the three centralities considered for $\pi^{\pm}\pi^{\pm}$, $\Kpm\Kpm$,
$\Kzs\Kzs$, $\pro\pro$, and $\pbar\pbar$. Statistical (thin lines) and systematic (boxes) uncertainties are shown. The $m_{\rm T}$ values for different centrality intervals are slightly
offset for clarity.}
\label{fig5}
\end{center}
\end{figure}

\begin{figure}
\begin{center}
\includegraphics[width=0.90\textwidth]{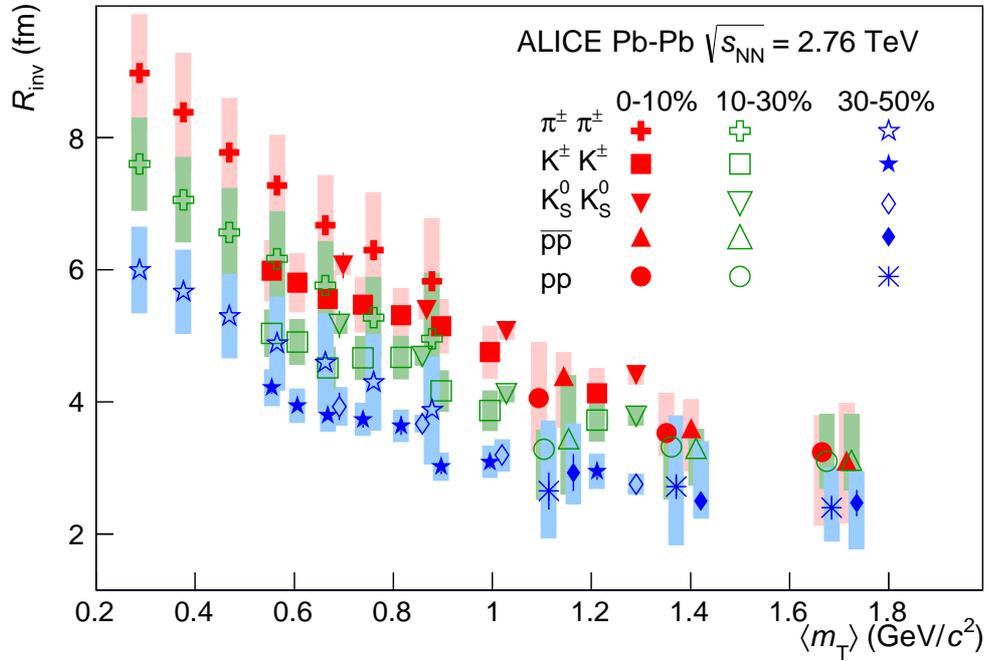}
\caption{(Color online) $R_{\rm inv}$ parameters vs. $m_{\rm T}$ for the three centralities considered for $\pi^{\pm}\pi^{\pm}$, $\Kpm\Kpm$,
$\Kzs\Kzs$, $\pro\pro$, and $\pbar\pbar$. Statistical (thin lines) and systematic (boxes) uncertainties are shown.}
\label{fig6}
\end{center}
\end{figure}

Figures \ref{fig5} and \ref{fig6} present the extracted fit parameters from $\pi^{\pm}\pi^{\pm}$, $\Kpm\Kpm$, $\Kzs\Kzs$, and $\pro\pro$ correlations for several intervals of centrality
and transverse mass. Both statistical and systematic uncertainties are shown. The quality of the fits used to extract the shown parameters can be assessed using the $\chi^2/NDF$ values, which are in the ranges of 1.2-5.0, 0.8-3.5, 0.6-1.5, and 0.8-3.2  for the pion, charged kaon, neutral kaon, and proton analyses, respectively.

Figure \ref{fig5} shows the extracted $\lambda$ parameters vs.\ $m_{\rm T}$ for several centralities. The proton $\lambda$ is the sum of $\lambda_{\rm pp}$ and $\lambda_{\rm p \Lambda}$ from Eq.~\ref{fitformula}. The values for all species
measured lie mostly in the range 0.3-0.7 
and show no significant centrality
dependence.
The values of $\lambda$ are less than unity due to long-lived resonances which dilute the correlation functions and also lead to non-Gaussian
shapes of the correlation functions, especially in the one-dimensional case~\cite{Abelev:2014pja}.
Results for kaons and protons are consistent with each other at similar $m_{\rm T}$. Values of $\lambda$ for pions are lower than for kaons due to the stronger influence of resonances; an additional cause could be a partial coherence of pions~\cite{Abelev:2013pqa}.

Figure ~\ref{fig6} shows the extracted $R_{\rm inv}$ parameters vs.\ $m_{\rm T}$ for several centralities. For overlapping $m_{\rm T}$, the radius parameters are mostly consistent with each other within uncertainties, though the pion radii are generally larger than the kaon radii.
The $\Kzs$ radii are slightly higher than $\Kpm$ radii for central collisions, but the difference is less than the systematic uncertainties. The radius parameters show
increasing size with increasing centrality as would be expected from a simple geometric picture of the
collisions. They also show a decreasing size with increasing $m_{\rm T}$ as would be expected
in the presence of collective radial flow~\cite{Akkelin:1995gh}. Both of these dependences can be seen in previous $\pi^{\pm}\pi^{\pm}$ femtoscopic
measurements~\cite{Lisa:2005dd,Aamodt:2011mr} and also reinforce the interpretation that collective flow
is present in these collisions for pions, kaons (neutral and charged), and protons alike.
Deviations from exact $m_{\rm T}$-scaling of $R_{\rm inv}$ can be explained as a consequence of the increase of the Lorentz factor with decreasing particle mass. In a hydrodynamic model~\cite{Kisiel:2014upa}, scaling is observed for the three-dimensional radii measured in the Longitudinally Co-Moving System (LCMS). The transformation from LCMS to PRF involves a boost along the outward direction only, where the boost value is proportional to the transverse velocity of the pair and inversely proportional to the particle mass (for similar $m_{\rm T}$). Thus, a smaller mass leads to an increase in the boosted $R_{\rm out}$ and, subsequently, $R_{\rm inv}$ in the PRF.
Indeed, we observe such an effect in the data, as pion radii are systematically higher than kaon radii at the same $m_{\rm T}$.

\begin{figure}
\begin{center}
\includegraphics[width=0.49\textwidth]{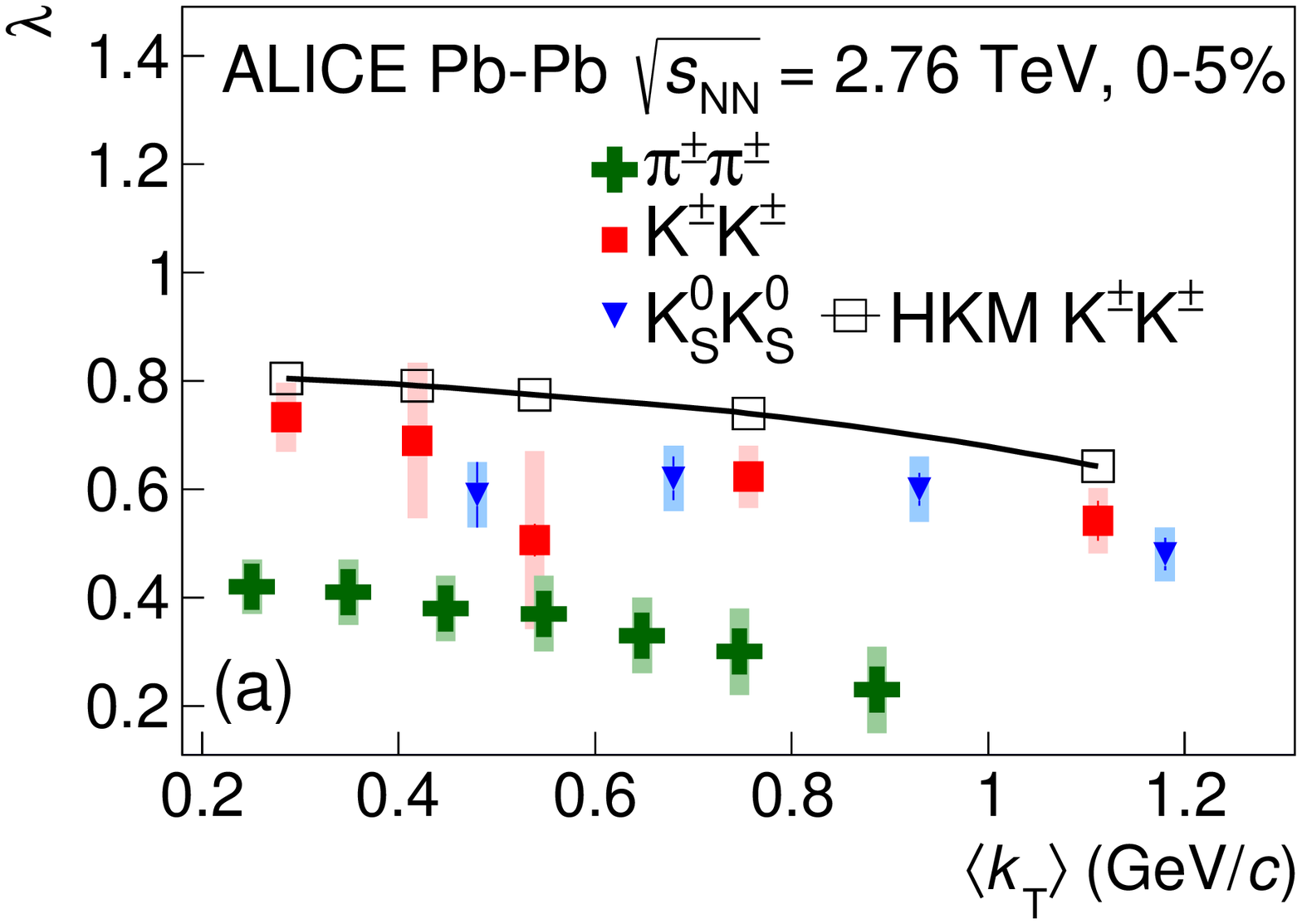}
\includegraphics[width=0.49\textwidth]{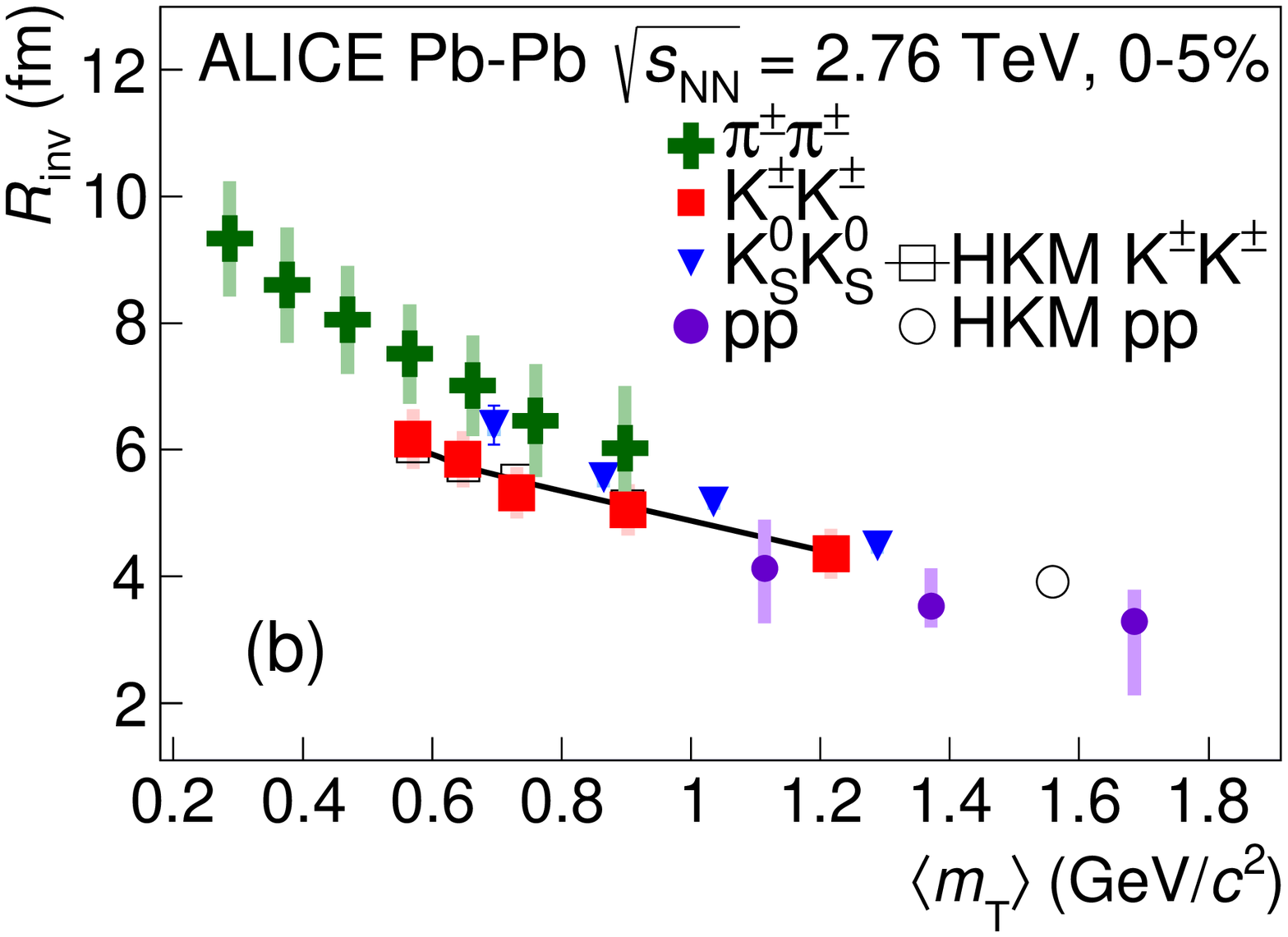}

\caption{(Color online) Comparison of the HKM model (see text)
with  measured kaon $\lambda$ (a) and $R_{\rm inv}$ (b) parameters for 0-5\% centrality. Statistical (thin lines) and systematic (boxes) uncertainties are shown.}
\label{fig7}
\end{center}
\end{figure}

A comparison of a hydrodynamic flow + kinetics model, HKM~\cite{Shapoval:2014wya},
with the measured $R_{\rm inv}$ and $\lambda$ parameters for 0-5\% centrality
is shown in Fig. \ref{fig7}. The HKM values in Fig. \ref{fig7} are specifically from $\Kpm\Kpm$, but the predictions for $\Kzs\Kzs$ and $\Kpm\Kpm$ are consistent
with each other.
For $R_{\rm inv}$, the charged kaon data show very good agreement with the predictions. The experimental data for the neutral kaons are again slightly higher than for the charged kaons, but this difference is still within systematic uncertainties.
For $\lambda$, both sets of kaon data match the decreasing trend with increasing $k_{\rm T}$ exhibited by the HKM points, but the model slightly overpredicts the data. It is shown in~\cite{Shapoval:2014wya} that the most important resonances for $\Kaon\Kaon$ pairs, $\Kaon$*(890) and $\phi$(1020), do not significantly influence the $\lambda$ parameter (due to their low contribution), and the discrepancy between the model and experimental data can be explained by the lower experimental kaon purity and deviations of the experimental correlation function shape from a Gaussian distribution. For protons, the HKM prediction is compatible with the data.
HKM calculations for one-dimensional pion radii are currently not available, but three-dimensional radii were reasonably reproduced by this model~\cite{Karpenko:2012yf}.

\section{Summary}
\label{sec:summary}
Results from femtoscopic studies of $\pi^{\pm}\pi^{\pm}$, $\Kpm\Kpm$, $\Kzs\Kzs$, $\pro\pro$, and $\pbar\pbar$ correlations from Pb-Pb collisions
at $\sqrt{s_{\mathrm {NN}}}=2.76$~TeV with ALICE at the LHC have been presented. The femtoscopic radii and $\lambda$ parameters were extracted from one-dimensional correlation
functions in terms of the invariant momentum difference. It was found that the emission source sizes of kaons and protons measured in these collisions exhibit
transverse mass scaling within uncertainties, which is consistent with hydrodynamic model predictions assuming collective flow. The deviation from the scaling for the pions can be explained 
as a consequence of the increase of the Lorentz factor with decreasing particle mass during the transformation from LCMS to PRF systems~\cite{Kisiel:2014upa}. The extracted $\lambda$ parameters are found to be less than unity, as
is expected due to long-lived resonances and non-Gaussian correlation functions. The predictions of the hydrokinetic model (HKM) for the one-dimensional femtoscopic radii for charged and neutral kaons and protons coincide well with the observations.

%
%

\newenvironment{acknowledgement}{\relax}{\relax}
\begin{acknowledgement}
\section*{Acknowledgements}

The ALICE Collaboration would like to thank all its engineers and technicians for their invaluable contributions to the construction of the experiment and the CERN accelerator teams for the outstanding performance of the LHC complex.
The ALICE Collaboration gratefully acknowledges the resources and support provided by all Grid centres and the Worldwide LHC Computing Grid (WLCG) collaboration.
The ALICE Collaboration acknowledges the following funding agencies for their support in building and
running the ALICE detector:
State Committee of Science,  World Federation of Scientists (WFS)
and Swiss Fonds Kidagan, Armenia,
Conselho Nacional de Desenvolvimento Cient\'{\i}fico e Tecnol\'{o}gico (CNPq), Financiadora de Estudos e Projetos (FINEP),
Funda\c{c}\~{a}o de Amparo \`{a} Pesquisa do Estado de S\~{a}o Paulo (FAPESP);
National Natural Science Foundation of China (NSFC), the Chinese Ministry of Education (CMOE)
and the Ministry of Science and Technology of China (MSTC);
Ministry of Education and Youth of the Czech Republic;
Danish Natural Science Research Council, the Carlsberg Foundation and the Danish National Research Foundation;
The European Research Council under the European Community's Seventh Framework Programme;
Helsinki Institute of Physics and the Academy of Finland;
French CNRS-IN2P3, the `Region Pays de Loire', `Region Alsace', `Region Auvergne' and CEA, France;
German Bundesministerium fur Bildung, Wissenschaft, Forschung und Technologie (BMBF) and the Helmholtz Association;
General Secretariat for Research and Technology, Ministry of
Development, Greece;
Hungarian Orszagos Tudomanyos Kutatasi Alappgrammok (OTKA) and National Office for Research and Technology (NKTH);
Department of Atomic Energy and Department of Science and Technology of the Government of India;
Istituto Nazionale di Fisica Nucleare (INFN) and Centro Fermi -
Museo Storico della Fisica e Centro Studi e Ricerche "Enrico
Fermi", Italy;
MEXT Grant-in-Aid for Specially Promoted Research, Ja\-pan;
Joint Institute for Nuclear Research, Dubna;
National Research Foundation of Korea (NRF);
Consejo Nacional de Cienca y Tecnologia (CONACYT), Direccion General de Asuntos del Personal Academico(DGAPA), M\'{e}xico, Amerique Latine Formation academique - European Commission~(ALFA-EC) and the EPLANET Program~(European Particle Physics Latin American Network);
Stichting voor Fundamenteel Onderzoek der Materie (FOM) and the Nederlandse Organisatie voor Wetenschappelijk Onderzoek (NWO), Netherlands;
Research Council of Norway (NFR);
National Science Centre, Poland;
Ministry of National Education/Institute for Atomic Physics and National Council of Scientific Research in Higher Education~(CNCSI-UEFISCDI), Romania;
Ministry of Education and Science of Russian Federation, Russian
Academy of Sciences, Russian Federal Agency of Atomic Energy,
Russian Federal Agency for Science and Innovations and The Russian
Foundation for Basic Research;
Ministry of Education of Slovakia;
Department of Science and Technology, South Africa;
Centro de Investigaciones Energeticas, Medioambientales y Tecnologicas (CIEMAT), E-Infrastructure shared between Europe and Latin America (EELA), Ministerio de Econom\'{i}a y Competitividad (MINECO) of Spain, Xunta de Galicia (Conseller\'{\i}a de Educaci\'{o}n),
Centro de Aplicaciones Tecnológicas y Desarrollo Nuclear (CEA\-DEN), Cubaenerg\'{\i}a, Cuba, and IAEA (International Atomic Energy Agency);
Swedish Research Council (VR) and Knut $\&$ Alice Wallenberg
Foundation (KAW);
Ukraine Ministry of Education and Science;
United Kingdom Science and Technology Facilities Council (STFC);
The United States Department of Energy, the United States National
Science Foundation, the State of Texas, and the State of Ohio;
Ministry of Science, Education and Sports of Croatia and  Unity through Knowledge Fund, Croatia;
Council of Scientific and Industrial Research (CSIR), New Delhi, India.
\end{acknowledgement}

\bibliographystyle{utphys}   
\bibliography{biblio}

\newpage
\appendix
\section{The ALICE Collaboration}
\label{app:collab}



\begingroup
\small
\begin{flushleft}
J.~Adam\Irefn{org40}\And
D.~Adamov\'{a}\Irefn{org83}\And
M.M.~Aggarwal\Irefn{org87}\And
G.~Aglieri Rinella\Irefn{org36}\And
M.~Agnello\Irefn{org111}\And
N.~Agrawal\Irefn{org48}\And
Z.~Ahammed\Irefn{org132}\And
S.U.~Ahn\Irefn{org68}\And
I.~Aimo\Irefn{org94}\textsuperscript{,}\Irefn{org111}\And
S.~Aiola\Irefn{org137}
% ...
\renewcommand\labelenumi{\textsuperscript{\theenumi}~}

\section*{Affiliation notes}
\renewcommand\theenumi{\roman{enumi}}
\begin{Authlist}
\item \Adef{0}Deceased
\item \Adef{idp23053484}{Also at: University of Kansas, Lawrence, Kansas, United States}
\end{Authlist}

\section*{Collaboration Institutes}
\renewcommand\theenumi{\arabic{enumi}~}
\begin{Authlist}

\item \Idef{org1}A.I. Alikhanyan National Science Laboratory (Yerevan Physics Institute) Foundation, Yerevan, Armenia
\item \Idef{org2}Benem\'{e}rita Universidad Aut\'{o}noma de Puebla, Puebla, Mexico
\item \Idef{org3}Bogolyubov Institute for Theoretical Physics, Kiev, Ukraine
\item \Idef{org4}Bose Institute, Department of Physics and Centre for Astroparticle Physics and Space Science (CAPSS), Kolkata, India
\item \Idef{org5}Budker Institute for Nuclear Physics, Novosibirsk, Russia
\item \Idef{org6}California Polytechnic State University, San Luis Obispo, California, United States
\item \Idef{org7}Central China Normal University, Wuhan, China
\item \Idef{org8}Centre de Calcul de l'IN2P3, Villeurbanne, France
\item \Idef{org9}Centro de Aplicaciones Tecnol\'{o}gicas y Desarrollo Nuclear (CEADEN), Havana, Cuba
\item \Idef{org10}Centro de Investigaciones Energ\'{e}ticas Medioambientales y Tecnol\'{o}gicas (CIEMAT), Madrid, Spain
\item \Idef{org11}Centro de Investigaci\'{o}n y de Estudios Avanzados (CINVESTAV), Mexico City and M\'{e}rida, Mexico
\item \Idef{org12}Centro Fermi - Museo Storico della Fisica e Centro Studi e Ricerche ``Enrico Fermi'', Rome, Italy
\item \Idef{org13}Chicago State University, Chicago, Illinois, USA
\item \Idef{org14}China Institute of Atomic Energy, Beijing, China
\item \Idef{org15}Commissariat \`{a} l'Energie Atomique, IRFU, Saclay, France
\item \Idef{org16}COMSATS Institute of Information Technology (CIIT), Islamabad, Pakistan
\item \Idef{org17}Departamento de F\'{\i}sica de Part\'{\i}culas and IGFAE, Universidad de Santiago de Compostela, Santiago de Compostela, Spain
\item \Idef{org18}Department of Physics and Technology, University of Bergen, Bergen, Norway
\item \Idef{org19}Department of Physics, Aligarh Muslim University, Aligarh, India
\item \Idef{org20}Department of Physics, Ohio State University, Columbus, Ohio, United States
\item \Idef{org21}Department of Physics, Sejong University, Seoul, South Korea
\item \Idef{org22}Department of Physics, University of Oslo, Oslo, Norway
\item \Idef{org23}Dipartimento di Elettrotecnica ed Elettronica del Politecnico, Bari, Italy
\item \Idef{org24}Dipartimento di Fisica dell'Universit\`{a} 'La Sapienza' and Sezione INFN Rome, Italy
\item \Idef{org25}Dipartimento di Fisica dell'Universit\`{a} and Sezione INFN, Cagliari, Italy
\item \Idef{org26}Dipartimento di Fisica dell'Universit\`{a} and Sezione INFN, Trieste, Italy
\item \Idef{org27}Dipartimento di Fisica dell'Universit\`{a} and Sezione INFN, Turin, Italy
\item \Idef{org28}Dipartimento di Fisica e Astronomia dell'Universit\`{a} and Sezione INFN, Bologna, Italy
\item \Idef{org29}Dipartimento di Fisica e Astronomia dell'Universit\`{a} and Sezione INFN, Catania, Italy
\item \Idef{org30}Dipartimento di Fisica e Astronomia dell'Universit\`{a} and Sezione INFN, Padova, Italy
\item \Idef{org31}Dipartimento di Fisica `E.R.~Caianiello' dell'Universit\`{a} and Gruppo Collegato INFN, Salerno, Italy
\item \Idef{org32}Dipartimento di Scienze e Innovazione Tecnologica dell'Universit\`{a} del  Piemonte Orientale and Gruppo Collegato INFN, Alessandria, Italy
\item \Idef{org33}Dipartimento Interateneo di Fisica `M.~Merlin' and Sezione INFN, Bari, Italy
\item \Idef{org34}Division of Experimental High Energy Physics, University of Lund, Lund, Sweden
\item \Idef{org35}Eberhard Karls Universit\"{a}t T\"{u}bingen, T\"{u}bingen, Germany
\item \Idef{org36}European Organization for Nuclear Research (CERN), Geneva, Switzerland
\item \Idef{org37}Excellence Cluster Universe, Technische Universit\"{a}t M\"{u}nchen, Munich, Germany
\item \Idef{org38}Faculty of Engineering, Bergen University College, Bergen, Norway
\item \Idef{org39}Faculty of Mathematics, Physics and Informatics, Comenius University, Bratislava, Slovakia
\item \Idef{org40}Faculty of Nuclear Sciences and Physical Engineering, Czech Technical University in Prague, Prague, Czech Republic
\item \Idef{org41}Faculty of Science, P.J.~\v{S}af\'{a}rik University, Ko\v{s}ice, Slovakia
\item \Idef{org42}Faculty of Technology, Buskerud and Vestfold University College, Vestfold, Norway
\item \Idef{org43}Frankfurt Institute for Advanced Studies, Johann Wolfgang Goethe-Universit\"{a}t Frankfurt, Frankfurt, Germany
\item \Idef{org44}Gangneung-Wonju National University, Gangneung, South Korea
\item \Idef{org45}Gauhati University, Department of Physics, Guwahati, India
\item \Idef{org46}Helsinki Institute of Physics (HIP), Helsinki, Finland
\item \Idef{org47}Hiroshima University, Hiroshima, Japan
\item \Idef{org48}Indian Institute of Technology Bombay (IIT), Mumbai, India
\item \Idef{org49}Indian Institute of Technology Indore, Indore (IITI), India
\item \Idef{org50}Inha University, Incheon, South Korea
\item \Idef{org51}Institut de Physique Nucl\'eaire d'Orsay (IPNO), Universit\'e Paris-Sud, CNRS-IN2P3, Orsay, France
\item \Idef{org52}Institut f\"{u}r Informatik, Johann Wolfgang Goethe-Universit\"{a}t Frankfurt, Frankfurt, Germany
\item \Idef{org53}Institut f\"{u}r Kernphysik, Johann Wolfgang Goethe-Universit\"{a}t Frankfurt, Frankfurt, Germany
\item \Idef{org54}Institut f\"{u}r Kernphysik, Westf\"{a}lische Wilhelms-Universit\"{a}t M\"{u}nster, M\"{u}nster, Germany
\item \Idef{org55}Institut Pluridisciplinaire Hubert Curien (IPHC), Universit\'{e} de Strasbourg, CNRS-IN2P3, Strasbourg, France
\item \Idef{org56}Institute for Nuclear Research, Academy of Sciences, Moscow, Russia
\item \Idef{org57}Institute for Subatomic Physics of Utrecht University, Utrecht, Netherlands
\item \Idef{org58}Institute for Theoretical and Experimental Physics, Moscow, Russia
\item \Idef{org59}Institute of Experimental Physics, Slovak Academy of Sciences, Ko\v{s}ice, Slovakia
\item \Idef{org60}Institute of Physics, Academy of Sciences of the Czech Republic, Prague, Czech Republic
\item \Idef{org61}Institute of Physics, Bhubaneswar, India
\item \Idef{org62}Institute of Space Science (ISS), Bucharest, Romania
\item \Idef{org63}Instituto de Ciencias Nucleares, Universidad Nacional Aut\'{o}noma de M\'{e}xico, Mexico City, Mexico
\item \Idef{org64}Instituto de F\'{\i}sica, Universidad Nacional Aut\'{o}noma de M\'{e}xico, Mexico City, Mexico
\item \Idef{org65}iThemba LABS, National Research Foundation, Somerset West, South Africa
\item \Idef{org66}Joint Institute for Nuclear Research (JINR), Dubna, Russia
\item \Idef{org67}Konkuk University, Seoul, South Korea
\item \Idef{org68}Korea Institute of Science and Technology Information, Daejeon, South Korea
\item \Idef{org69}KTO Karatay University, Konya, Turkey
\item \Idef{org70}Laboratoire de Physique Corpusculaire (LPC), Clermont Universit\'{e}, Universit\'{e} Blaise Pascal, CNRS--IN2P3, Clermont-Ferrand, France
\item \Idef{org71}Laboratoire de Physique Subatomique et de Cosmologie, Universit\'{e} Grenoble-Alpes, CNRS-IN2P3, Grenoble, France
\item \Idef{org72}Laboratori Nazionali di Frascati, INFN, Frascati, Italy
\item \Idef{org73}Laboratori Nazionali di Legnaro, INFN, Legnaro, Italy
\item \Idef{org74}Lawrence Berkeley National Laboratory, Berkeley, California, United States
\item \Idef{org75}Lawrence Livermore National Laboratory, Livermore, California, United States
\item \Idef{org76}Moscow Engineering Physics Institute, Moscow, Russia
\item \Idef{org77}National Centre for Nuclear Studies, Warsaw, Poland
\item \Idef{org78}National Institute for Physics and Nuclear Engineering, Bucharest, Romania
\item \Idef{org79}National Institute of Science Education and Research, Bhubaneswar, India
\item \Idef{org80}Niels Bohr Institute, University of Copenhagen, Copenhagen, Denmark
\item \Idef{org81}Nikhef, Nationaal instituut voor subatomaire fysica, Amsterdam, Netherlands
\item \Idef{org82}Nuclear Physics Group, STFC Daresbury Laboratory, Daresbury, United Kingdom
\item \Idef{org83}Nuclear Physics Institute, Academy of Sciences of the Czech Republic, \v{R}e\v{z} u Prahy, Czech Republic
\item \Idef{org84}Oak Ridge National Laboratory, Oak Ridge, Tennessee, United States
\item \Idef{org85}Petersburg Nuclear Physics Institute, Gatchina, Russia
\item \Idef{org86}Physics Department, Creighton University, Omaha, Nebraska, United States
\item \Idef{org87}Physics Department, Panjab University, Chandigarh, India
\item \Idef{org88}Physics Department, University of Athens, Athens, Greece
\item \Idef{org89}Physics Department, University of Cape Town, Cape Town, South Africa
\item \Idef{org90}Physics Department, University of Jammu, Jammu, India
\item \Idef{org91}Physics Department, University of Rajasthan, Jaipur, India
\item \Idef{org92}Physik Department, Technische Universit\"{a}t M\"{u}nchen, Munich, Germany
\item \Idef{org93}Physikalisches Institut, Ruprecht-Karls-Universit\"{a}t Heidelberg, Heidelberg, Germany
\item \Idef{org94}Politecnico di Torino, Turin, Italy
\item \Idef{org95}Purdue University, West Lafayette, Indiana, United States
\item \Idef{org96}Pusan National University, Pusan, South Korea
\item \Idef{org97}Research Division and ExtreMe Matter Institute EMMI, GSI Helmholtzzentrum f\"ur Schwerionenforschung, Darmstadt, Germany
\item \Idef{org98}Rudjer Bo\v{s}kovi\'{c} Institute, Zagreb, Croatia
\item \Idef{org99}Russian Federal Nuclear Center (VNIIEF), Sarov, Russia
\item \Idef{org100}Russian Research Centre Kurchatov Institute, Moscow, Russia
\item \Idef{org101}Saha Institute of Nuclear Physics, Kolkata, India
\item \Idef{org102}School of Physics and Astronomy, University of Birmingham, Birmingham, United Kingdom
\item \Idef{org103}Secci\'{o}n F\'{\i}sica, Departamento de Ciencias, Pontificia Universidad Cat\'{o}lica del Per\'{u}, Lima, Peru
\item \Idef{org104}Sezione INFN, Bari, Italy
\item \Idef{org105}Sezione INFN, Bologna, Italy
\item \Idef{org106}Sezione INFN, Cagliari, Italy
\item \Idef{org107}Sezione INFN, Catania, Italy
\item \Idef{org108}Sezione INFN, Padova, Italy
\item \Idef{org109}Sezione INFN, Rome, Italy
\item \Idef{org110}Sezione INFN, Trieste, Italy
\item \Idef{org111}Sezione INFN, Turin, Italy
\item \Idef{org112}SSC IHEP of NRC Kurchatov institute, Protvino, Russia
\item \Idef{org113}SUBATECH, Ecole des Mines de Nantes, Universit\'{e} de Nantes, CNRS-IN2P3, Nantes, France
\item \Idef{org114}Suranaree University of Technology, Nakhon Ratchasima, Thailand
\item \Idef{org115}Technical University of Ko\v{s}ice, Ko\v{s}ice, Slovakia
\item \Idef{org116}Technical University of Split FESB, Split, Croatia
\item \Idef{org117}The Henryk Niewodniczanski Institute of Nuclear Physics, Polish Academy of Sciences, Cracow, Poland
\item \Idef{org118}The University of Texas at Austin, Physics Department, Austin, Texas, USA
\item \Idef{org119}Universidad Aut\'{o}noma de Sinaloa, Culiac\'{a}n, Mexico
\item \Idef{org120}Universidade de S\~{a}o Paulo (USP), S\~{a}o Paulo, Brazil
\item \Idef{org121}Universidade Estadual de Campinas (UNICAMP), Campinas, Brazil
\item \Idef{org122}University of Houston, Houston, Texas, United States
\item \Idef{org123}University of Jyv\"{a}skyl\"{a}, Jyv\"{a}skyl\"{a}, Finland
\item \Idef{org124}University of Liverpool, Liverpool, United Kingdom
\item \Idef{org125}University of Tennessee, Knoxville, Tennessee, United States
\item \Idef{org126}University of the Witwatersrand, Johannesburg, South Africa
\item \Idef{org127}University of Tokyo, Tokyo, Japan
\item \Idef{org128}University of Tsukuba, Tsukuba, Japan
\item \Idef{org129}University of Zagreb, Zagreb, Croatia
\item \Idef{org130}Universit\'{e} de Lyon, Universit\'{e} Lyon 1, CNRS/IN2P3, IPN-Lyon, Villeurbanne, France
\item \Idef{org131}V.~Fock Institute for Physics, St. Petersburg State University, St. Petersburg, Russia
\item \Idef{org132}Variable Energy Cyclotron Centre, Kolkata, India
\item \Idef{org133}Vin\v{c}a Institute of Nuclear Sciences, Belgrade, Serbia
\item \Idef{org134}Warsaw University of Technology, Warsaw, Poland
\item \Idef{org135}Wayne State University, Detroit, Michigan, United States
\item \Idef{org136}Wigner Research Centre for Physics, Hungarian Academy of Sciences, Budapest, Hungary
\item \Idef{org137}Yale University, New Haven, Connecticut, United States
\item \Idef{org138}Yonsei University, Seoul, South Korea
\item \Idef{org139}Zentrum f\"{u}r Technologietransfer und Telekommunikation (ZTT), Fachhochschule Worms, Worms, Germany
\end{Authlist}
\endgroup


\end{document}